




\magnification\magstep1 
\baselineskip13pt
\vsize23.6truecm


\input miniltx
\input graphicx





\def\hatt{\widehat}

\def\half{\hbox{$1\over2$}}

\def\normal{{\cal N}}
\def\sumin{\sum_{i=1}^n}

\def\RR{\mathord{I\kern-.3em R}}
\def\PP{\mathord{I\kern-.3em P}}
\def\NN{\mathord{I\kern-.3em N}}
\def\ZZ{\mathord{I\kern-.3em Z}} 
\def\Var{{\rm Var}}

\def\mtrix{\pmatrix} 

\def\subsection{\medskip}

\font\bigbf=cmbx12

\font\csc=cmcsc10

 at 10truept 
\font\smallrm=cmr8 
\font\smallsl=cmsl8

\font\smallbf=cmbx8

\def\today{\number\day \space \ifcase\month\or
January\or February\or March\or April\or May\or June\or 
July\or August\or September\or October\or November\or December\fi  
\space \number\year}


   
\def\ref#1{{\noindent\hangafter=1\hangindent=20pt
  #1\smallskip}}          

\def\quotationone{\smallrm Where there is a Will}
\def\quotationtwo{\smallrm There is a Won't}
\def\hskipdistanceleft{\hskip-3.5pt}
\def\hskipdistanceright{\hskip-2.0pt}
\footline={{
\ifodd\count0
        {\hskipdistanceleft\quotationone\phantom{\smallrm\today}
                \hfil{\rm\the\pageno}\hfil
         \phantom{\quotationone}{\smallrm\today}\hskipdistanceright}
        \else 
        {\hskipdistanceleft\quotationtwo\phantom{\today}
                \hfil{\rm\the\pageno}\hfil
         \phantom{\quotationtwo}{\smallrm\today}\hskipdistanceright}
        \fi}}

         
\def\cstok#1{\leavevmode\thinspace\hbox{\vrule\vtop{\vbox{\hrule\kern1pt
        \hbox{\vphantom{\tt/}\thinspace{\tt#1}\thinspace}}
        \kern1pt\hrule}\vrule}\thinspace} 
 


\def\fermat#1{\setbox0=\vtop{\hsize4.00pc
        \smallrm\raggedright\noindent\baselineskip9pt
        \rightskip=0.5pc plus 1.5pc #1}\leavevmode
        \vadjust{\dimen0=\dp0
        \kern-\ht0\hbox{\kern-4.00pc\box0}\kern-\dimen0}}

\def\hsizeplusepsilon{14.25truecm} 
\def\fermatright#1{\setbox0=\vtop{\hsize4.00pc
        \smallrm\raggedright\noindent\baselineskip9pt
        \rightskip=0.5pc plus 1.5pc #1}\leavevmode
        \vadjust{\dimen0=\dp0
        \kern-\ht0\hbox{\kern\hsizeplusepsilon\box0}\kern-\dimen0}}



\def\hskipdistanceleft{\hskip-3.5pt}
\def\hskipdistanceright{\hskip-2.0pt}

\def\today{November 1994}
 
\def\quotationone{\smallrm Nils Lid Hjort}
\def\quotationtwo{\smallrm The Olympic 500 meter}

\centerline{\bigbf Should the Olympic sprint skaters run the 500 meter twice?}

\smallskip
\centerline{\bigbf Nils Lid Hjort$^*$\footnote{}
        {\smallsl\baselineskip10pt\hskip-20pt $^*$
        Hjort is Professor of Statistics at the University of Oslo,
        Research Statistician with the Norwegian Computing Centre, and 
        Member of the World's Speed Skating Statisticians' Association.
        \vskip-0.30truecm}}


{{\medskip\narrower\noindent\baselineskip11pt
{\csc Abstract.} 
The Olympic 500 meter sprint competition 
is the `Formula One event' of speed skating, 
and is watched by millions of television viewers. 
A draw decides who should start in inner lane 
and who in outer lane. Many skaters dread the last inner lane,
where they need to tackle heavier centrifugal forces 
than their companions in the last outer lane, 
at maximum speed around 55 km/hour, at a time when fatigue may set in. 
The aim of this article is to investigate this potential
difference between last inner and last outer lane. 
For this purpose data from eleven Sprint World Championships 
1984--1994 are exploited. 
A bivariate mixed effects model is used that in addition to 
the inner-outer lane information takes account of different 
ice and weather conditions on different days, 
unequal levels for different skaters, and the passing times 
for the first 100 meter. 
The underlying `unfairness parameter', estimated with optimal precision, 
is about 0.05 seconds, and is indeed significantly different from zero; 
it is about three times as large as its estimated standard deviation. 

\smallskip\noindent 
Results from the work reported on here played a decisive role in 
leading the International Skating Union and the 
International Olympic Committee to change the rules for the 500 meter 
sprint event; as of the Nagano 1998 Olympic Games, the sprinters 
are to skate twice, with one start in inner lane and one in outer lane. 
The best average result determines the final list, 
and the best skaters from the first run 
are paired to skate last in the second run. 
It has also been decided that the same rules shall apply 
for the single distance 500 meter World Championships;
these are arranged yearly from 1996 onwards.  

\smallskip\noindent\sl  
{\csc Key words and phrases:}
combining data sources, 
Dan Jansen, 
mixed effects model, 
Olympic Games, 
speed skating, 
Sprint World Championships, 
unfairness parameter 
\smallskip}}

\bigskip
\centerline{\bf 1. Background: Is the Olympic 500 meter unfair?}

\medskip\noindent
{\sl ``He drew lane with anxious attentiveness --- and could not 
conceal his disappointment: First outer lane! 
He threw his blue cap on the ice with a resigned movement, 
but quickly contained himself and picked it up again. 
With a start in inner lane he could have set a world record,
perhaps be the first man in the world under 40 seconds.
Now the record was hanging by a thin thread --- possibly 
the gold medal too. At any rate he couldn't tolerate any further mishaps.''} 

This book excerpt (from Bj\o rnsen, 1963, Ch.~9) 
illustrates what has been known for a long time, 
that even the most accomplished sprint skaters 
experience difficulties with the last inner lane. 
(Yevgeni Grishin indeed had his famous technical accident 
there and found himself stumbling into outer lane,
caused by the leather on his left boot touching the ice
as he had to lean over at the high speed; 
he miraculously went on to win the 1960 Squaw Valley Olympic gold medal 
at 40.2 seconds. Some days later he achieved 39.6, 
this time with last outer lane.)
The last inner lane skaters have to fight a higher acceleration force, 
since the inner track radius is about 25--26 meter and 
the outer track about 29--30 meter,
at a time when speed is at peak and fatigue may set in; see Figure 1.  
The acceleration force formula is $mv^2/r$, where $m$ is mass, 
$v$ is velocity, and $r$ is radius. 
Thus a 90 kg skater with top speed 
400 meter by 27 seconds 
meets a force of about 80 kp in inner lane and about 70 kp in outer lane. 
%
As a consequence many skaters 
are not able to keep to the designated curve, 
glide out towards or even into the outer lane, 
and in such cases have to skate some extra distance. 
This phenomenon is particularly prominent 
on rinks where the ice is fast and the curvature radius minimal,
as for the modern indoor rinks.   
The last inner lane skaters are also more prone to 
experiencing technical accidents. 


\bigskip


\centerline{\includegraphics[scale=0.66]{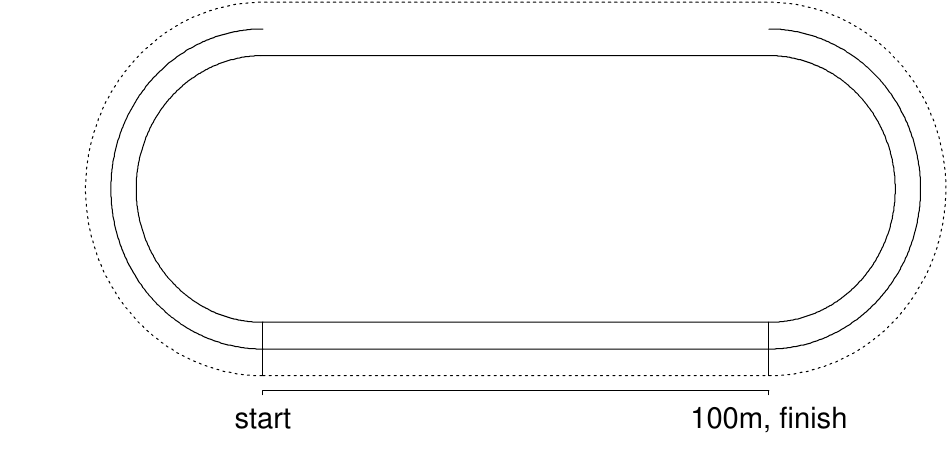}}



{\medskip\sl\narrower\noindent\baselineskip12pt
{\csc Figure 1}.
{\sl The speed skating rink. The skaters exchange lanes 
on the backstraight. One lap, comprising one outer and 
one inner lane, is exactly 400 meter long.}  
\smallskip} 


\subsection
{\csc 1.1. The present investigation.} 
This problem has been recognised and solved 
in a satisfactory manner since 1975 when it comes 
to the annual Sprint World Championships events,
in that an International Skating Union rule was enforced 
to make skaters have last inner and last outer lanes on alternate days. 
But the Olympic event has a potential unfairness built into it 
since the skaters run only once,
and half of them are allotted last inner lane and the other half 
last outer lane, by chance. 

My aim has been to estimate the potential `unfairness' difference 
parameter in question as precisely as possible,
and to test whether the unfairness is or is not statistically significant. 
The data I have used consist of the complete 500 meter lists 
from the eleven Sprint World Championships for Men 
Trondheim 1984--Calgary 1994. Each skater has a 100 meter passing time 
and a 500 meter result for the Saturday event, 
and similarly for the Sunday event, 
and skaters start in inner and outer tracks on alternative days. 
(The SWCs also include 1000 meter races on both Saturday and Sunday,
and all four runs contribute to the final standing; 
our present concern lies however with the 500 meter only.) 

Agree to define $d$ as the average difference between 
a result reached by last inner track versus last outer track,
for the average top sprinter: 
$$d=\hbox{average difference 
        (last inner track -- last outer track)}. \eqno(1.1)$$ 
This is the statistical parameter we wish to estimate 
with optimal precision. 
It is easy to put up some simple estimates based on 
individual differences, for example, but such estimates 
will typically lack in statistical precision. 
I have used a more sophisticated statistical approach, 
involving a model with ingredients 
$$\eqalign{
{\rm 500\,m}_{\rm day\,1}&=a_1+b_1\cdot{\rm 100\,m}_{\rm day\,1} 
        +c_{\rm skater} \pm\half d+{\rm variation}_{\rm day\,1}, \cr 
{\rm 500\,m}_{\rm day\,2}&=a_2+b_2\cdot{\rm 100\,m}_{\rm day\,2} 
        +c_{\rm skater} \pm\half d+{\rm variation}_{\rm day\,2}. \cr}
                        \eqno(1.2)$$
The $\pm$ here is a plus if the skater has last inner track 
and a minus if the skater has last outer track. 
The model and methods used take account of the 100 meter passing times,
the individual skater's ability, 
possible differences in day-to-day conditions, 
and the inner-outer information. 

The model is more formally motivated and described in Section 2,
and estimation methods along with analysis of their precision 
are developed in Section 3. 
I find it reasonable to interpret $d$ in a somewhat conservative manner,
and do not want it to be overly influenced by various extreme 
or too unusual results; it is meant to be the average difference 
for `usual runs' where the skaters are up to their normal best.  
It is therefore necessary to decide on criteria to define outliers,
run times that are subsequently to be removed from the final analysis. 
Three outlier tests are given in Section 4, which also contains 
material and analysis that support the fundamental statistical model used.  

{{\smallskip\baselineskip11pt\narrower
\def\hf{\hfill}
\def\ha{\hskip1.50cm} 
\def\hb{\hskip1.25cm} 
\def\hc{\phantom{-}}
\settabs\+\ha\hf 1989 &Heerenveen: \hb&--0.019 \hskip0.50cm &0.086 \cr 
\+\hf      &            &$\hc\hc \hatt d$  &$\!\!$st.~error \cr 

\smallskip
\+\hf 1994 &Calgary:    &$\hc 0.010$  &0.043 \cr 
\+\hf 1993 &Ikaho:      &$\hc 0.032$  &0.041 \cr  
\+\hf 1992 &Oslo:       &$  - 0.019$  &0.086 \cr   
\+\hf 1991 &Inzell:     &$\hc 0.023$  &0.040 \cr  
\+\hf 1990 &Troms\o:    &$\hc 0.096$  &0.087 \cr 
\+\hf 1989 &Heerenveen: &$\hc 0.128$  &0.047 \cr 
\+\hf 1988 &West Allis: &$   -0.147$  &0.090 \cr 
\+\hf 1987 &Sainte Foy: &$   -0.151$  &0.080 \cr
\+\hf 1986 &Karuizawa:  &$\hc 0.035$  &0.066 \cr  
\+\hf 1985 &Heerenveen: &$\hc 0.090$  &0.058 \cr
\+\hf 1984 &Trondheim:  &$\hc 0.131$  &0.038 \cr  
\+\hf      &grand average: &$\hc 0.048$ &0.016 \cr 

\medskip\noindent
{\csc Table 1.} 
{\sl Estimates of the difference parameter $d$ for eleven 
Sprint World Championships for men, along with standard errors 
(estimated standard deviation) for these.}
\smallskip}}

\subsection
{\csc 1.2. Conclusions.} 
In view of the potentially serious implications 
(as measured on the Olympic scale) of the results 
and their interpretations I decided to gather enough data to 
reach a standard deviation of the final $d$ estimate of about 0.02 seconds. 
Preliminary analysis based on a couple of SWCs indicated that 
this meant including about ten complete SWC data sets. 
Section 5 reports more fully on analysis of the eleven SWCs 1984--1994;
presently we exhibit an estimate of the $d$ parameter for each,
together with their standard errors (estimated standard deviations)
and their grand (weighted) average. 

The $d$ estimates are seen to vary somewhat from competition to competition, 
as was expected in view of the sometimes quite different conditions
regarding wind, temperature, gliding conditions, and the actual 
sizes of the inner and outer lane radii. In situations where there
is a strong wind against the skater on the backstraight, 
`the crossing stretch', as arguably was the case for many competitors 
in Sainte Foy 1987, West Allis 1988 and Oslo 1992, for example, 
it may actually be fortuitous to have the last inner course. 
The $d$ estimates are indeed found to be negative,
and the corresponding standard errors highest,
for exactly these three events. 
The $d$s for other events have values in the range 0.09--0.13
(Trondheim 1984, Heerenveen outdoor 1985, 
Heerenveen indoor 1989, Troms\o{} 1990),
and in the more modest range 0.01--0.04 
(Karuizawa 1986, Inzell 1991, Ikaho 1993 and Calgary 1994).
It is a visible feature of the data that the more 
statistically secure $d$ estimates are those that are high,
and typically reflect good and stable conditions with runners 
attaining their normal high standards.    
As a case in point, if one excludes 
1987 Sainte Foy, 1988 West Allis, 1990 Troms\o{} and 1992 Oslo, 
which are the four events with standard errors for $d$ estimates 
exceeding 0.08, then the grand average estimate for the remaining 
seven SWCs is 0.065, which is nearly four times its standard error 0.017
(95\% confidence interval $[0.032,0.098]$). 
The grand (weighted) average estimate for all eleven SWCs is 0.048 seconds, 
which is about three times its standard error 0.016,
and hence statistically significant 
($p$-value 0.001; 95\% confidence interval $[0.174,0.079]$). 
Further details, along with supplementary analysis 
and graphical evidence supporting the $d>0$ conclusion, 
are given in Section 5. 

Folklore knowledge in the stands has it that women are not 
as bothered with the last inner lane as men sometimes are; 
they are seen to keep much better to the designated curve, 
and also seem to have fewer slips and technical accidents than men. 
This is supported by data; see the discussion of Section 6. 
Thus the unfairness argument is much weaker in case of the 
women's event. 

How important is the 0.05 seconds difference? 
At top speed 26 seconds for 400 meters a skater 
manages an impressive 15.4 meter in 1 second, 
and about 0.75 meter in 0.05 seconds, 
which translates into roughly 0.15 percent. 
It would mean about 15 meters in a 10 000 m run, 
and about 65 meters in a marathon. So the difference matters! 
This is also borne out through comparison with real and 
simulated result lists for the 1994, 1992, 1988 Olympics, 
given in Appendix I. 
As alluded to above there are also reasons to believe that the real 
$d$ number is {\it larger} for the modern indoor rinks that 
will host the future Olympic speed skating events, 
than the grand average value 0.05 arrived at here. 

\smallskip
{\csc 1.3. A proposal to the ISU and the IOC.} 
A natural proposal to the International Skating Union and the 
International Olympic Committee, in view of these findings, 
is that the skaters should run the 500 meter twice,
with one start in inner and one in outer lane,
as in the Sprint World Championships. 
The most natural solution would then be as in alpine events 
and ski jumping, with a `reversed starting list', 
with the best skaters from the first run starting the latest in 
the second run, subject to correct pairing 
with respect to inner and outer lane, and with the best average
result defining the final ranking. 
This is a spectator-friendly regime too; 
large screens would at every stage inform viewers of the 
current ranking as well as the performances 
required by the next skaters to reach the top of the list.  
The average viewers would then easier 
understand when to be appropriately excited. 
\eject 

\subsection
{\csc 1.4. Coda: They said yes.} 
Such a proposal was put forward to the ISU at their June 1992 congress 
in Davos, along with a brief summary of the work presented here
(based on SWCs 1984--1992), through 
the Norwegian delegation. 
Also included as handout material to sell the argument 
were real and simulated 500 meter result lists from the 
two last Olympic events, where the simulated list in question was 
the the speculated outcome if the inner lane starters had started 
in outer lane and vice versa, and computed under the $d=0.06$ assumption 
(which was the estimate based on 1984--1992 data;
see Appendix I below for such lists for the 1988, 1992, 1994 events,
using the $d=0.05$ figure).
It is fair to add that the attraction argument,
that the proposed scenario would actually be (even) 
more spellbinding for spectators, was emphasised as much as the 
statistical questions related to the unfairness parameter.
The meeting decided not to interfere with the already laid plans 
for the 1994 Lillehammer Olympics, and to reconsider the matter 
at their next congress in June 1994 in Boston. 
And at this meeting the representatives of the 34 attending countries 
(ISU members) after some debate unanimously voted yes to the new proposal, 
which was this time put forward by the ISU Technical Committee 
for Speed Skating, to be made effective from the 1998 Nagano Olympics onwards,
for both men and women. The new 500 meter rule is also to be 
made effective for the annual World Championships for single distances,
which are introduced as from 1996.  

\bigskip
\centerline{\bf 2. Sprint World Championships data and the statistical model}

\medskip\noindent 
This section describes the data and motivates the statistical model 
which is used. A simpler model for 
the differences of finishing times is also considered.  

\def\hf{\hfill}
\def\ha{\hskip0.22cm} 
\def\hb{\phantom{1}} 
\def\hc{\hskip1.33cm}
\def\in{\hskip0.04cm i}

{{\smallskip\baselineskip11pt\narrower 
\settabs\+\hc&\hf 24. &Shakshakbayev \hb
        &o\ha &10.16\ha &38.49\quad &o\ha &10.02\ha &37.91 \cr 
\medskip
\+\hc&\hf     & 
        & &{\bf first day:} & & &{\bf second day:} \cr 
\+\hc&\hf     & 
        & &{\bf 100} &{\bf 500} & &{\bf 100} &{\bf 500} \cr 
\smallskip        
\+\hc&\hf 1. &D.~Jansen             
        &o &{\hb}9.82 &35.96    &\in &{\hb}9.75 &35.76 \cr 
\+\hc&\hf 2. &S.~Klevchena          
        &o &{\hb}9.78 &36.39    &\in &{\hb}9.82 &36.27 \cr
\+\hc&\hf 3. &J.~Inoue              
        &\in &{\hb}9.98 &36.43  &o &{\hb}9.76 &36.05 \cr
\+\hc&\hf 4. &H.~Shimizu            
        &o &{\hb}9.70 &36.35    &\in &{\hb}9.77 &36.08 \cr
\+\hc&\hf 5. &K.~Scott              
        &\in &{\hb}9.96 &36.87  &o &{\hb}9.87 &36.55 \cr
\+\hc&\hf 6. &I.~Zhelezovsky        
        &\in &10.08 &36.90      &o &10.22 &36.99 \cr
\+\hc&\hf 7. &T.~Kuroiwa            
        &\in &10.11 &36.83      &o &10.07 &36.75 \cr
\+\hc&\hf 8. &Y.-M.~Kim             
        &o &{\hb}9.81 &36.56    &\in &{\hb}9.76 &36.53 \cr
\+\hc&\hf     & ... and so on ... \cr  

\medskip\noindent 
{\csc Table 2.}
{\sl Results for the best of the 33 skaters taking part in the 
Sprint World Championship in Calgary, Candada, January 29--30 1994. 
Here `i' and `o' signify that the skater started in respectively 
outer and inner lane.} 
\smallskip}}

\subsection
{\csc 2.1. The bivariate mixed effects model.} 
For illustration, consider data$^*$\footnote{}
        {\smallsl\baselineskip10pt\hskip-20pt $^*$
        Data source: {\csc sk\o ytenytt} (`Speed Skating News'), 
        5/6/1984, 
        6/1985,
        6/1986,
        6/1987,
        7 and 8/1988,
        6/1989,
        5/1990, 
        7/1991,
        7/1992,
        5/1993, 
        3 and 6/1994.  
        This is an international bulletin issued by the 
        World Speed Skating Statisticians' Association about ten times a year,
        and with contributing associate editors from about 20 countries.
        Chief editor is Magne Teigen, Veggli, Norway. 
        \vskip-0.30truecm}
from the SWC 1994, held in the Olympic Oval, Calgary, Canada. 
These are of the form given in Table 2; 
a full listing of the 1984--1994 results is offered in Appendix II.

\noindent
We represent the inner-outer information as 
$$z_{1,i}=\cases{-1 &if no.~$i$ starts in inner track on day 1, \cr 
       \phantom{-}1 &if he starts in outer track on day 1, \cr} $$
with a similar $z_{2,i}$ for day 2. 
Note that $z_{2,i}$ is always $-z_{1,i}$, 
by the ISU rules for these Championships. 
Let furthermore $x_{1,i}$ and $Y_{1,i}$ be 100 meter time and 
finishing 500 meter time for skater $i$ on day 1, 
and similarly $x_{2,i}$ and $Y_{2,i}$ are 100 meter time and 
finishing 500 meter result for the same skater on day 2.

A natural starting point for the model building procedure 
is the representation 
$$\eqalign{
Y_{1,i}&=a_1+b_1x_{1,i}+c_i+\half d z_{1,i}+e_{1,i}, \cr 
Y_{2,i}&=a_2+b_2x_{2,i}+c_i+\half d z_{2,i}+e_{2,i} \cr}\eqno(2.1)$$
for $i=1,\ldots,n$, 
where $d$ is the quantity of primary interest, see (1.1);
a skater needs the extra amount $\half d$ if he starts in outer 
lane compared to the extra amount $-\half d$ if he starts in inner lane. 
Further, $e_{i,1}$ and $e_{i,2}$ represent random statistical
variation for the two runs, modelled as $2n$ independent 
terms distributed as $\normal\{0,\sigma^2\}$, 
while $c_i$ represents that particular skater's ability
compared to the average level in the competition. 
Thus (2.1) is the formal version of the model that was hinted at in (1.2). 
Top skaters Dan Jansen and Igor Zhelezovsky have negative $c_i$s, 
perhaps around $-1$, while those found in the second half or so 
on the final results list have positive $c_i$s, 
perhaps around $1$ for the slower ones.  
The $c_i$s are not directly observable; 
they would be poorly estimated based on data from one competition,
and although more accurate values for these could be assigned
based on extensive data from several competitions, 
the intention presently is to treat them in the `random effects' way.
Assuming $c_1,\ldots,c_n$ to come from a normal $\{0,\kappa^2\}$, 
the model takes the form 
$$\mtrix{Y_{1,i} \cr Y_{2,i} \cr}
  \sim\normal_2\{\mtrix{a_1+b_1x_{1,i}+dw_i \cr a_2+b_2x_{2,i}-dw_i \cr},
  \mtrix{\sigma^2+\kappa^2 &\kappa^2 \cr 
        \kappa^2 &\sigma^2+\kappa^2 \cr}\}. \eqno(2.2)$$
Here, for convenience, 
$$w_i=\half z_{1,i}=-\half z_{2,i}
=\cases{\phantom{-}1/2&if outer start on day 1 
                and inner start on day 2, \cr 
                  -1/2&if inner start on day 1 
                and outer start on day 2. \cr} \eqno(2.3)$$
The `intraclass correlation' $\rho=\kappa^2/(\sigma^2+\kappa^2)$ parameter 
is here representing the stability for the average skater. 

The Saturday parameters $(a_1,b_1)$ and Sunday parameters $(a_2,b_2)$
are a priori different since gliding and other conditions, like 
wind and temperature, are often different for the two days. 
There is an argument favouring $b_1=b_2$, though, 
which means that the day-to-day difference in overall conditions
can be well explained by the difference $a_2-a_1$ alone. 
Suppose for example that the conditions are a bit worse on Sunday.
This leads to somewhat larger $x_2$ values than $x_1$ values, 
and somewhat larger $Y_2$ values than $Y_1$ values, by about the same factor;
the two slopes in question would be approximately the same.
In the analysis I therefore put $b_1=b_2$ in (2.2). 
Data support this assumption, see Section 4. 
This reduction from five to four mean parameters is not overly important;
results based using all five parameters give very nearly the same results.
I have similarly tested whether it is necessary to have 
different $\sigma_1$ and $\sigma_2$ parameters in (2.1)--(2.2),
but data again have supported the simpler model with $\sigma_1=\sigma_2$. 
Another suggestion, in the present quest to employ as
few parameters as naturally possible, is to let 
$a_2=ka_1$, $b_2=kb_1$, and $\sigma_2=k\sigma_1$,
with a $k$ factor envisaged to be quite close to 1. 
The $b_1=b_2$ parameterisation is however easier and more effective.  

\subsection
{\csc 2.2. A simpler model based on differences.}
There is a rather simple alternative way of obtaining
an estimate of $d$, based on the observed differences; 
$$\eqalign{
Y_{2,i}-Y_{1,i}
&=a_2-a_1+b(x_{2,i}-x_{1,i})+d(\half z_{2,i}-\half z_{1,i})
        +\eta_{2,i}-\eta_{1.i} \cr 
&\sim\normal\{a_2-a_1+b(x_{2,i}-x_{1,i})-2dw_i,2\sigma^2\}. \cr}\eqno(2.4)$$
Here the individual $c_i$ capability parameters disappear,
and one circumvents the need for binormal analysis; 
ordinary linear regression gives $\hatt d_{\rm simple}$, say,
in addition to $\hatt a_{0,{\rm simple}}$, $\hatt b_{\rm simple}$, 
and $\hatt\sigma_{\rm simple}$, where $a_0=a_2-a_1$. 
The $d$ estimate based on all $n$ pairs will be slightly more 
precise, however, as discussed in Section 3.3 below. 
To carry out outlier testing and a part of the 
final overall analysis, to be discussed in Section 5,
it is necessary also to estimate $\kappa$, 
on which the (2.4) differences throw no light. 
In addition it is of course preferable
to have as precise estimates of all parameters involved 
as possible, and $b$, in particular, is more precisely estimated 
in the full model than in the simpler differences model. 
We also note that the binormal (2.2) model, with $b_1=b_2=b$, 
is equivalent to stochastic independence between 
difference $Y_{2,i}-Y_{1,i}$ and average $\bar Y_i=(Y_{1,i}+Y_{2,i})/2$,
and with 
$$\eqalign{
\bar Y_i&\sim\normal\{\bar a+b\bar x_i,\kappa^2+\sigma^2/2\}, 
        \quad {\rm where\ }\bar x_i=(x_{1,i}+x_{2,i})/2, \cr
Y_{2,i}-Y_{1,i}&\sim\normal\{a_2-a_1+b(x_{2,i}-x_{1,i})-2dw_i,
                2\sigma^2\}. \cr} \eqno(2.5)$$ 

\bigskip
\centerline{\bf 3. Parameter estimates and their precision} 

\medskip\noindent 
In this section the parameter estimation procedure 
for models of the type (2.2) is outlined.
We also include analysis of the precision of the parameters, 
caring particularly about the $d$ estimators.  

\subsection
{\csc 3.1. Estimation in the mixed effects model.}
Suppose in general terms that we have $n$ independent pairs of data
$$\eqalign{Y_{1,i}&=x_{1,i}'\beta+f_{1,i}
                   =x_{1,i,1}\beta_1+\cdots+x_{1,i,p}\beta_p+f_{1,i}, \cr
           Y_{2,i}&=x_{2,i}'\beta+f_{2,i}
                   =x_{2,i,1}\beta_1+\cdots+x_{2,i,p}\beta_p+f_{2,i}, \cr}$$
where $\beta=(\beta_1,\ldots,\beta_p)'$ 
is the vector of regression coefficients, 
$x_{1,i}$ is a $p$-covariate vector for $Y_{1,i}$, 
and $x_{2,i}$ is a $p$-covariate vector for $Y_{2,i}$. 
Furthermore, $(f_{1,i},f_{2,i})'$ is zero mean binormal 
with covariance $\kappa^2$ and variances $\sigma^2+\kappa^2$;
in the notation above $f_{1,i}=c_i+e_{1,i}$ and $f_{2,i}=c_i+e_{2,i}$. 
In other words, 
$$\mtrix{Y_{1,i}\cr Y_{2,i} \cr}
        \sim\normal_2\{\mtrix{x_{i,1}'\beta \cr x_{i,2}'\beta \cr}, 
        \pmatrix{\sigma^2+\kappa^2 &\kappa^2 \cr
                          \kappa^2 &\sigma^2+\kappa^2}
        ={\sigma^2\over 1-\rho}
         \pmatrix{1 &\rho \cr \rho &1 \cr}\}, $$
the latter matrix expression being in terms 
of the intraclass correlation parameter $\rho=\kappa^2/(\sigma^2+\kappa^2)$. 

Aiming to find the maximum likelihood (ML) estimators, 
we start with the log-likelihood for data. It is 
$$\log L(\beta,\sigma,\rho)
        =-2n\log\sigma-{1\over2}n\log{1+\rho\over 1-\rho}
        -{1\over2}{1\over \sigma^2}{Q(\beta)\over 1+\rho}, \eqno(3.1)$$
where 
$$\eqalign{
Q(\beta)&=\sumin (Y_i-x_i'\beta)'\pmatrix{1 &-\rho \cr -\rho &1}
        (Y_i-x_i'\beta) \cr
        &=\sumin (Y_{1,i}-x_{1,i}'\beta)^2
         +\sumin (Y_{2,i}-x_{2,i}'\beta)^2
         -2\rho\sumin (Y_{1,i}-x_{1,i}'\beta)(Y_{2,i}-x_{2,i}'\beta) \cr
        &=Q_1(\beta)+Q_2(\beta)-2\rho Q_3(\beta). \cr}$$
We are to find the maximisers $\hatt\beta$, $\hatt\sigma$, $\hatt\rho$
of (3.1). The one for $\beta$ must be $\hatt\beta=\hatt\beta(\hatt\rho)$,
where $\hatt\beta(\rho)$ is found from minimisation of 
$Q(\beta)$, i.e.
$$\hatt\beta(\rho)=\{M_{11}+M_{22}-\rho(M_{12}+M_{21})\}^{-1}
        \{S_{11}+S_{22}-\rho(S_{12}+S_{21})\}, \eqno(3.2)$$
in which 
$$M_{uv}={1\over n}\sumin x_{u,i}x_{v,i}' 
        \quad {\rm and} \quad 
  S_{uv}={1\over n}\sumin x_{u,i}Y_{v,i}, \quad u,v=1,2.$$ 
And the ML estimators for $\sigma$ and $\rho$ are found from 
maximising $\log L(\hatt\beta(\rho),\sigma,\rho)$. 
Taking partial derivatives w.r.t.~$\sigma$ and $\rho$ 
gives two equations that must be obeyed: 
$$\hatt\sigma^2(\rho)={1\over 1+\rho} 
                {Q(\hatt\beta(\rho))\over 2n}
        \quad {\rm and} \quad 
  \hatt\sigma^2(\rho)={1-\rho \over 1+\rho}
{Q_1(\hatt\beta(\rho))+Q_2(\hatt\beta(\rho))+2Q_3(\hatt\beta(\rho)) \over 2n}. 
                        \eqno(3.3)$$
Finding the ML estimator $\hatt\rho$ is accomplished 
either via maximisation of the log-likelihood profile function 
$$\eqalign{
\log L(\hatt\beta(\rho),\hatt\sigma(\rho),\rho)
        &=n\Bigl[-\log\Bigl\{ {1\over 1+\rho}
                {Q(\hatt\beta(\rho))\over 2n}\Bigr\}
         +{1\over 2}\log{1-\rho \over 1+\rho}-1\Bigr] \cr
        &=n\bigl[\half\log(1-\rho^2)
         -\log\{Q(\hatt\beta(\rho))/2n\}-1\bigr],\cr} \eqno(3.4)$$
or from solving for the two expressions in (3.3) to be equal.
Note also that 
$$\rho={2Q_3(\hatt\beta(\rho))\over 
        Q_1(\hatt\beta(\rho))+Q_2(\hatt\beta(\rho))} \eqno(3.5)$$
at the parameter point that solves (3.3), i.e.~for the ML solution. 
This fits the fact that
$Q_1(\beta)$, $Q_2(\beta)$, $Q_3(\beta)$
have expected values $\sigma^2/(1-\rho)$, $\sigma^2/(1-\rho)$, 
and $\sigma^2\rho/(1-\rho)$, at the true model. 
Solving (3.5) for $\rho$ is yet another method for carrying out 
the ML estimation. 

We will actually use a sample-size modification for $\hatt\sigma$.
Assuming for a moment that the value of $\rho$ is known, 
one shows easily that $\hatt\beta(\rho)$ of (3.2) is normal, unbiased, 
and with covariance matrix $n^{-1}\sigma^2(1+\rho)M_\rho^{-1}$, where 
$$M_\rho=M_{11}+M_{22}-\rho(M_{12}+M_{21}). $$
Furthermore arguments can be furnished, 
using orthogonalisation techniques and properties of the binormal distribution,
to demonstrate that 
$$2n\,\hatt\sigma^2(\rho)/(1+\rho)
        =Q(\hatt\beta(\rho))/(1+\rho)\sim 2\sigma^2\chi^2_{2n-p}, $$
and that $Q(\hatt\beta(\rho))$ 
is statistically independent of $\hatt\beta(\rho)$. 
In particular this invites using the modified estimator  
$$\hatt\sigma^2_{\rm un}
        ={1\over 1+\hatt\rho}
        {Q(\hatt\beta(\hatt\rho))\over 2n-p}
        ={2n\over 2n-p}\hatt\sigma^2\,; \eqno(3.6)$$
it is slightly larger than the ML estimate $\hatt\sigma^2$, 
to take estimation variability of the $p$ regression coefficients into account.
We will also similarly use 
$\hatt\kappa_{\rm un}^2=\hatt\sigma_{\rm un}^2\hatt\rho/(1-\hatt\rho)$
as a sample-size corrected version of the ML estimator for $\kappa^2$. 
The argument is not disturbed by the insertion of $\hatt\rho$ 
for $\rho$, since we show in a minute that   
$\hatt\rho$ is approximately independent of 
$\hatt\beta=\hatt\beta(\hatt\rho)$. 

\subsection
{\csc 3.2. Precision of estimates.}
Note that classic regression theory distributional results do not hold here,
for example, $\hatt\beta$ is not quite normally distributed 
since the random $\hatt\rho$ is inserted. 
But traditional distributional approximations for ML estimators 
can be appealed to and implies that 
$(\hatt\beta,\hatt\sigma,\hatt\rho)$ is approximately 
jointly normally distributed with the correct mean vector 
and covariance matrix $J^{-1}/n$, where 
$J$ is minus the mean of the normalised 
and twice differentiated log-likelihood,
calculated at the true parameter values. 
With some efforts one finds 
$$J=\pmatrix{
        M_\rho/\{\sigma^2(1+\rho)\} &0 &0 \cr
        0 &4/\sigma^2 & 2/\{\sigma(1-\rho^2)\} \cr
        0 &2/\{\sigma(1-\rho^2)\} &2/(1-\rho^2)^2 \cr}. $$
This means that 
$$\hatt\beta\approx\normal_p\{\beta,
        n^{-1}\,\sigma^2(1+\rho)M_\rho^{-1}\}, \eqno(3.7)$$
that 
$$\pmatrix{\hatt\sigma \cr \hatt\rho}
        \approx\normal_2\bigl\{\pmatrix{\sigma \cr \rho \cr},\,
        {1\over n}\pmatrix{\half \sigma^2 &-\half\sigma(1-\rho^2) \cr 
                  -\half\sigma(1-\rho^2) &(1-\rho^2)^2 \cr}\bigr\},$$
and that $\hatt\beta$ and $(\hatt\sigma,\hatt\rho)$ 
become independent for large $n$.  
Confidence intervals and tests can now be furnished in the usual fashion.
We comment specifically on this below for the case of $d$.

The (3.7) result continues to be a very good approximation
even if the exact conditions of the normal model (2.2) should 
be violated. This follows from standard large-sample theory.
The distributional results for $\hatt\sigma$ and $\hatt\rho$
might have to be adjusted for nonnormality of $(Y_{1,i},Y_{2,i})$;
the more generally correct approximate variance of $\hatt\sigma$ 
can for example be shown to be $n^{-1}\sigma^2(\half+{1\over4}{\rm kurt})$,
where ${\rm kurt}$ is the kurtosis of $Y_{1,i}-Y_{2,i}$. 
In any case this would not have seriously hampered our analysis,
which is concerned primarily with $d$ and the other mean parameters,
but it is comforting that in fact no serious departure from normality
could be detected in our data (as soon as the quite few 
outliers were removed), as further commented on in Section 4. 

\def\ave{{\rm ave}}

\subsection
{\csc 3.3. Comparison of two $d$ estimators.} 
The binormal mixed effects model used is that of (2.2), 
with mean parameter vector $(a_1,a_2,b,d)'$ and with 
covariate vectors $(1,0,x_{1,i},w_i)'$ and $(0,1,x_{2,i},-w_i)'$
associated with skater $i$. 
The (3.7) result is that $(\hatt a_1,\hatt a_2,\hatt b,\hatt d)$ 
has covariance matrix approximately equal to 
$n^{-1}\sigma^2(1+\rho)M_\rho^{-1}$. 
In this situation, 
$$\eqalign{
M_{11}&=\mtrix{1         &0 &\ave(x_1)    &\ave(w) \cr 
                0         &0 &0            &0       \cr  
                \ave(x_1) &0 &\ave(x_1^2)  &\ave(x_1w) \cr 
                \ave(w)   &0 &\ave(x_1)    &\ave(w^2) \cr}, \cr 
M_{22}&=\mtrix{0         &0         &0            &0 \cr 
                0         &1         &\ave(x_2)    &\ave(-w)   \cr  
                0         &\ave(x_2) &\ave(x_2^2)  &\ave(-x_2w) \cr 
                0         &\ave(-w)  &\ave(-x_2w)  &\ave(w^2) \cr}, \cr
M_{12}&=\mtrix{0         &1         &\ave(x_2)    &\ave(-w) \cr 
                0         &0         &0            &0   \cr  
                0         &\ave(x_1) &\ave(x_1x_2) &\ave(-x_1w) \cr 
                0         &\ave(w)   &\ave(x_2w)   &\ave(-w^2) \cr}, \cr 
M_{21}&=\mtrix{0         &0         &0            &0  \cr 
                1         &0         &\ave(x_1)    &\ave(w)   \cr  
                \ave(x_2) &0         &\ave(x_1x_2) &\ave(x_2w) \cr 
                \ave(-w)  &0         &\ave(-x_1w)  &\ave(-w^2) \cr}, \cr}$$
where $\ave(x_1)$ is the average $n^{-1}\sumin x_{1,i}$, and so on. 
These can be computed for the given covariates $x_1$, $x_2$, $w$,
and this is what is used to produce standard errors 
(estimated standard deviations) for $\hatt d$ in the SWCs;
see Section 4 and Table 1 of Section 1. 

The statistical analysis takes place conditional 
on covariates $x_1$, $x_2$, $w$, 
but it is also illuminating to study simpler approximations, 
based on `typical behaviour' of these,
whose values also can also be viewed as outcomes of random mechanisms. 
In such a framework it is clear that $w$ behaves 
stochastically independent of $x_1$ and $x_2$,
and that the symmetric $\pm\half$ variable $w$ has mean zero;
in fact $\ave(w)=0$ if $n$ is even and $\pm\half/n$ if $n$ is odd.
It follows that $\ave(x_1w)$ and $\ave(x_2w)$ will both be close to zero. 
Using these arguments it follows that 
$$M_\rho\approx\mtrix{1                 &-\rho &\ave(x_1-\rho x_2)     &0 \cr
           -\rho             &1     &\ave(x_2-\rho x_1)     &0 \cr
           \ave(x_1-\rho x_2)&\ave(x_2-\rho x_1) 
                        &\ave(x_1^2+x_2^2-2\rho x_1x_2)     &0 \cr 
           0                 &0     &0          &\half+\half\rho \cr}. $$      
In particular this demonstrates that 
$$\Var\,\hatt d\approx n^{-1}\sigma^2(1+\rho)\,2/(1+\rho)
        =2\sigma^2/n. \eqno(3.8)$$

The simpler estimate $\hatt d_{\rm simple}$ 
based on differences, as in (2.4), can be analysed similarly. 
It emerges by ordinary linear regression of $Y_{2,i}-Y_{1,i}$
against $x_{2,i}-x_{1,i}$ and $-2w_i$. The covariance matrix 
for say $(\hatt a_{0,\rm simple},\hatt b_{\rm simple},\hatt d_{\rm simple})$
constructed in this fashion, where $a_0=a_2-a_1$, is equal to 
$$\eqalign{
&{2\sigma^2\over n}
        \mtrix{1            &\ave(x_2-x_1)     &\ave(-2w) \cr 
               \ave(x_2-x_1)&\ave((x_2-x_1)^2) &\ave(-2w(x_2-x_1)) \cr 
               \ave(-2w)    &\ave(-2w(x_2-x_1))&\ave(4w^2) \cr}^{-1} \cr 
&\qquad\qquad  
 \approx {2\sigma^2\over n}
        \mtrix{1            &\ave(x_2-x_1)     &0  \cr 
               \ave(x_2-x_1)&\ave((x_2-x_1)^2) &0  \cr 
               0            &0                 &1  \cr}^{-1}. \cr}$$
It follows that $\hatt d_{\rm simple}$ has approximately 
the very same precision as the more complicated $\hatt d$ that builds
more explicitly on the binormal mixed effects model. 
These are approximations, however, and the $\hatt d$ will have
a slight edge in each application, since the differences-based 
simpler method builds on less information. 

We also note that the $b$ parameter will typically 
be better estimated in the mixed model than in the differences model. 
Calculations as above show that $\hatt b$ has $(\sigma^2/2n)(1+\rho)/\tau^2$ 
while $\hatt b_{\rm simple}$ has $(\sigma^2/\tau^2)/n$ 
as approximate variances, where $\tau^2$ is empirical variance of $x$s.
Thus the ratio $\Var\,\hatt b_{\rm simple}/\Var\,\hatt b$ is $2/(1+\rho)$. 
In the same vein it can be shown that $\hatt\sigma_{\rm un}$ 
and the $\hatt\sigma_{\rm simple}$, available by regression on (2.4),
have approximately the same precision.

\bigskip
\centerline{\bf 4. Deciding on outliers, and validating the model}

\medskip\noindent 
As mentioned in the introduction we would not wish to 
see $d$ overly influenced by unusual results. 
Minor slips or accidents occur frequently in this 
technically demanding sport, and easily cause 
losses of tenths of a second; we envisage our $d$ as the average 
difference for normal runs without such mishaps. 
This is one of several reasons favouring a robust analysis of the model 
and in particular a statistically robust estimate of $d$. 
While various robust procedures are available, I opt for 
the conceptually simple method of detecting outliers first 
and removing these from the final analysis. This is also
quite reasonable in view of the fact that data convincingly 
support the parametric model (2.2), as discussed below.

\subsection
{\csc 4.1. Outlier criteria.} 
Let $r_{1,i}$ and $r_{2,i}$ the the lap times for the skater's 
last 400 meter. These can be expressed as 
$$\eqalign{
r_{1,i}=Y_{1,i}-x_{1,i}&=a_1+(b-1)x_{1,i}+dw_i+c_i+e_{1,i}, \cr
r_{2,i}=Y_{2,i}-x_{2,i}&=a_2+(b-1)x_{2,i}-dw_i+c_i+e_{1,i}, \cr} \eqno(4.1)$$
It is natural to discard cases where one of these are too large,
in comparison with expected normal behaviour, 
and also cases where the absolute difference $|r_{2,i}-r_{1,i}|$ 
between the skater's two lap times is too large.
This statistical safety net will catch skaters who have had 
minor or not so minor technical slips in one of his or her two runs. 

\centerline{\includegraphics[scale=0.66]{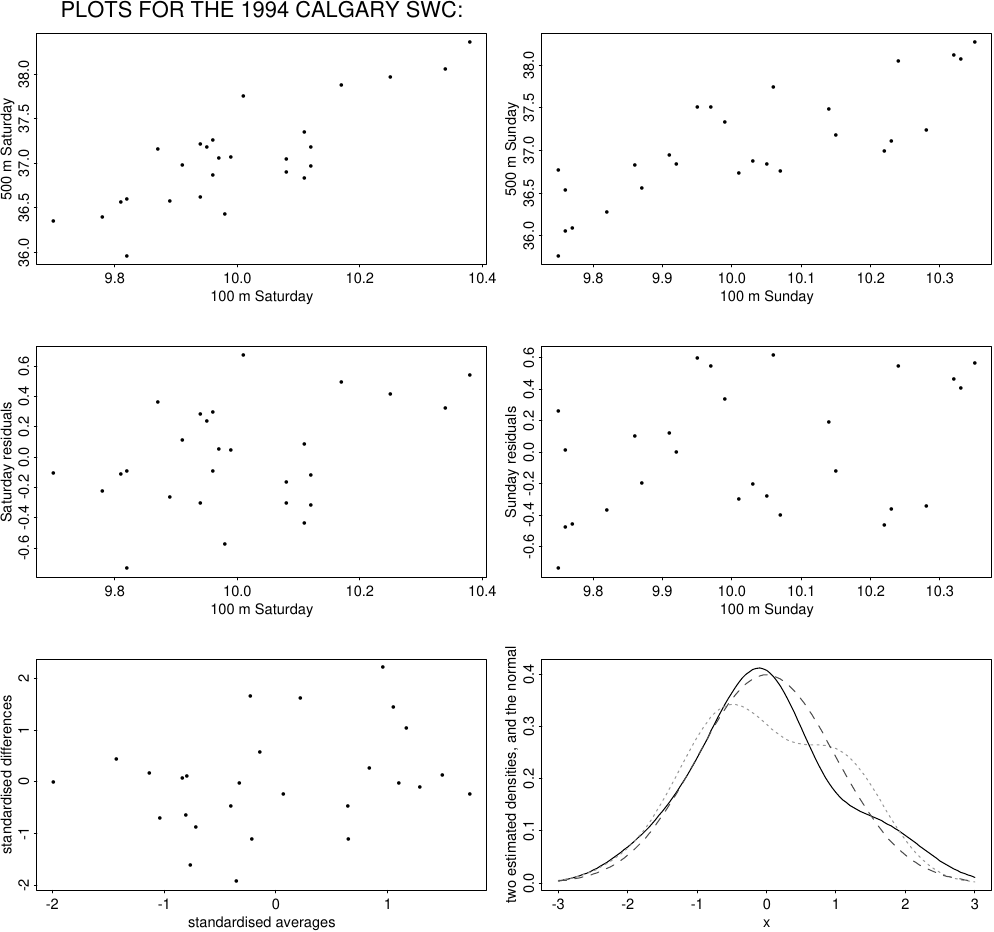}}


{{\medskip\narrower\baselineskip11pt\noindent 
{\csc Figure 2} (a)--(f). 
{\sl Plots to validate the mixed effects statistical model,
using results from the 1994 Calgary Sprint World Championship. 
Plots (a) and (b) give $Y_i$ versus $x_i$ for each day 
and show acceptable linearity 
(as well as Dan Jansen's World Record 35.76 in plot (b)
and Hiroyasu Shimizu's 9.70 in plot(a)). 
Plots (c) and (d) display Saturday and Sunday residuals 
against 100 meter passing times, 
and support the assumption of constant variability across $x$ values.  
In (e) standardised differences ${\rm diff}^*_i$ are plotted 
against standardised averages ${\rm ave}^*_i$, see (4.4), 
with non-significant correlation 0.249. 
Finally (f) gives nonparametrically 
estimated densities of the ${\rm diff}^*_i$s (fully drawn curve)
and of the ${\rm ave}^*_i$s (dotted curve), using the traditional 
kernel method, together with the standard normal curve.}
\smallskip}}

The first outlier criterion is developed as follows. 
Based on all data from a two-day SWC event, 
excluding only those who fell or were disqualified, 
estimates are computed of $(a_1,a_2,b,d)$ and of $(\sigma,\kappa)$. 
Then we form the outlier test statistics 
$$\eqalign{
t_{1,i}=\{Y_{1,i}-(\hatt a_1+\hatt bx_{1,i}+\half\hatt dz_{1,i})\}
                /(\hatt\sigma_{\rm un}^2+\hatt\kappa_{\rm un}^2)^{1/2}, \cr 
t_{2,i}=\{Y_{2,i}-(\hatt a_2+\hatt bx_{2,i}+\half\hatt dz_{2,i})\}
                /(\hatt\sigma_{\rm un}^2+\hatt\kappa^2_{\rm un})^{1/2}, \cr} 
                        \eqno(4.2)$$ 
both of which are approximately a standard normal if the skater 
had a `normal' run without accidents. 
Slightly more accurate denominators could be constructed here,
dividing by an estimate of the exact standard deviation rather 
than with the approximate one, 
but the difference is unsubstantial for the present purposes. 
We deem a case `above normal bounds' if $t_{1,1}$ or $t_{2,i}$ exceed 2.75. 
Note that we allow cases with unusually {\it low} values of $t$,
since these correspond to extraordinary {\it good} performances. 

The second outlier criterion emerges by looking at normal 
behaviour of the difference between lap times. 
By (4.1) this leads to constructing 
$$\eqalign{t_{3,i}
&=\bigl[r_{2,i}-r_{1,i}-\{\hatt a_2-\hatt a_1
        +(\hatt b-1)(x_{2,i}-x_{1,i})-2\hatt d w_i\}\bigr]
        /\sqrt{2}\hatt\sigma_{\rm un} \cr 
&=\{(Y_{2,i}-\hatt a_2-\hatt b x_{2,i}-\hatt d w_i)
    -(Y_{1,i}-\hatt a_1-\hatt b x_{2,i}+\hatt d w_i)\}
        /\sqrt{2}\hatt\sigma_{\rm un}. \cr} \eqno(4.3)$$
Again these should be approximately distributed as standard normals 
if the skater's two runs were `normal'.
A case is deemed `outside normal bounds' if $|t_{3,i}|\ge2.75$. 

\subsection
{\csc 4.2. Model validation.} 
The adequacy of the basic model was checked against various 
potential violations, and, in short, 
no audible objections were raised by the data. 

Plots of $Y_i$ versus $x_i$ for each day, 
and for each competition,
agreed well with the assumed linearity. 
Plots of Saturday and Sunday residuals against 
respectively $x_{1,i}$ and $x_{2,i}$ revealed no departure 
from the assumed constant level of variability hypothesis. 
The simultaneous aspects of the (2.2) model were checked 
via the reformulated equivalent (2.5) model. 
Scatters were plotted and correlations computed to check for
possible dependencies between averages and differences, 
and again these supported the mean and 
variance/covariance structure of the model. 
See Figure 2 (a)--(e) which illustrate these features 
for the case of the 1994 Calgary Championship. 

Coming finally to checking the hypothesised bi-Gau\ss ian distribution, 
this could be separated into one-dimensional
normality assessment of averages and differences, as with (2.5).  
We have pointed out already, in Section 3.2, 
that deviations from normality is of no great 
concern as far as the $d$ analysis is concerned,
and that such deviations, specifically in the form of 
kurtosis values different from the zero predicted by normality, 
at most could cause mild concern for precision
of the $\sigma$ and $\kappa$ estimates. 
However, the normal distribution fits nicely. 
For purposes of plotting and for comparison over different 
Championships it is convenient to standardise these, as 
$$\eqalign{
{\rm ave}^*_i&=\{\bar Y_i-(\hatt{\bar a}+\hatt b\bar x_i)\}
        /(\hatt\kappa_{\rm un}^2+\hatt\sigma^2_{\rm un}/2)^{1/2}, \cr 
{\rm diff}^*_i&=\{Y_{2,i}-Y_{1,i}-(\hatt a_2-\hatt a_1
        +\hatt b(x_{2,i}-x_{1,i})-2\hatt d w_i)\}
        /\sqrt{2}\hatt\sigma_{\rm un}. \cr} \eqno(4.4)$$
These have zero mean and variance nearly equal to 1,
and under the model hypothesis these should behave independently 
of each other and each be approximately standard normal. 
Figure 2(e) displays the 28 pairs of these variables 
for the 1994 SWC in Calgary (of the 33 participants, 
three had falls, and two were declared outliers, by
the criteria above; see Appendix II), 
and Figure 2(f) shows smooth estimates 
of their densities, comparing them also with the standard normal. 
The fit is quite acceptable. The same was observed on the basis 
of quantile-quantile plots against the standard normal. 
The skewnesses were 0.324 and 0.269, 
and the kurtosis values were $-0.149$ and $-0.703$, 
all values well within normal range. 
(Approximate 90\% ranges for the skewness and kurtosis,
with 28 normal data points, are approximately $\pm0.76$ and $\pm1.52$.)   
The observed correlation is 0.249 and is also within the 
90\% expected range (which is approximately $\pm0.31$). 
Similar plots were made and coefficients computed 
for the other Championships as well, and again no challenges
to the model assumptions were made. 

\subsection
{\csc 4.3. How many SWCs should I analyse?} 
The present investigation is a statistical detective search
for a quite tiny parameter, in a way looking for the odd meter in 500
with a statistician's magnifying glass. 
We cannot expect to be able to declare a positive $d$ with 
data from only one or two Championships. 
There is accordingly a question of gathering enough
data to create a reasonable statistical power for the envisioned 
size of $d$, say about 0.05 seconds. From (3.8) it emerges 
that the optimally combined estimator $\hatt d_{\rm grand}$, 
taken over say $K$ SWCs, would have a variance of the form 
$\{\sum_{j=1}^Kn_j/(2\sigma_j^2)\}^{-1}$, with $n_j$ skaters
in the $j$th SWC. As an approximation this is the same as 
$2\sigma^2/N$, where $N=\sum_{j=1}^Kn_j$, and analysis 
from the first couple of SWCs data sets I used suggested 
that $\sigma\approx0.25$. As a rough guide, therefore,
about 300 skaters' paired runs were necessary to gather, 
in order to form a grand estimate with standard error of size 0.02 seconds
(95\% confidence interval width around 0.08 seconds). 
Since around 30 skaters compete in each SWC I needed to go through
about ten complete SWCs to achieve the necessary precision.
A standard error of 0.02 seconds would give detection probability 
around 80\% for a true $d=0.05$, and around 90\% for $d=0.06$.

\bigskip
\centerline{\bf 5. Analysing the Sprint World Championships 1984--1994} 

\medskip\noindent 
This section summarises analysis carried out for each of the 
eleven Championships 1984--1994, using methods developed 
in Section 3. Only skaters who passed all three outlier tests above
were included. 

First we give a table of all required parameter estimates,
for each of the eleven situations. The $d$ estimate column is identical
to the list also given in Table 1 of Section~1, 
where also standard errors were given. 
Interpretation of the parameters is discussed in Sections 2 and 3. 
Again we note that the few cases where the $d$ estimate actually 
is negative are also characterised by higher than normal values
for $\sigma$, in particular, which means higher variability 
around each skater's normal level. On the occasions 
in question a tentative explanation lies with the 
partly severe and variable weather conditions that met 
the participants (Sainte Foy 1987, West Allis 1988, Oslo 1992). 
The cases with more than a tiny difference between $a_1$ and $a_2$,
like for Sainte Foy 1987 and Troms\o{} 1990,  
are the events where Saturday and Sunday conditions differed markedly. 

The primary method for testing $d=0$ is simply to consider 
the natural overall estimate and compare its value to 
its standard error. If $\hatt d_j$ is the estimate in year $j$, 
with standard error ${\rm se}_j$, then 
$$\hatt d=\sum_{j=1984}^{1994}\hatt d_j/{\rm se}_j^2
        \Big/\sum_{j=1984}^{1994}1/{\rm se}_j^2 \eqno(5.1)$$
is the optimal combination, with precision given by 
$\Var\,\hatt d=(\sum_{j=1984}^{1994}1/{\rm se}_j^2)^{-1}$. 
(This variance calculation is exactly valid in the case of 
fully known values for the ${\rm se}_j$s; 
that these are estimated rather than fully known does however only 
cause a very modest second order level increase in the variance.)
These are the formulae that produced the grand estimate 0.048
with standard error 0.016 in Table 1 of Section 1. 
Again, this is significant with $p$-value 0.001. 

{{\smallskip\baselineskip11pt\narrower
\def\hf{\hfill}
\def\ha{\quad} 
\def\hb{\hskip1.25cm} 
\def\hc{\phantom{-}}
\def\hd{\hskip0.60cm}
\def\he{\phantom{1}}
\settabs\+\ha\hf 1994: \hd&17.485\hd &17.442\hd &1.961\hd &0.009\hd\hd  
        &0.882\hd &0.152\hd &0.415 \cr   
\smallskip

\+\hf       &$\hc a_1$   &$\hc a_2$   &$\hc b$    &$\hc\hc d$  
        &$\hc\rho$ &$\hc\sigma$ &$\hc\kappa$\cr
\+{\hf}1994: &16.984 &16.938 &2.007 &$\hc0.010$ &0.838  &0.156  &0.355 \cr   
\+{\hf}1993: &18.861  &18.998  &1.892  &$\hc0.032$ &0.799  &0.158  &0.315 \cr
\+{\hf}1992: &13.541  &13.369  &2.494  &$-0.019$   &0.538  &0.308  &0.332 \cr
\+{\hf}1991: &13.184  &13.060  &2.462  &$\hc0.023$ &0.800  &0.155  &0.309 \cr
\+{\hf}1990: &15.137  &14.552  &2.371  &$\hc0.096$ &0.338  &0.327  &0.234 \cr
\+{\hf}1989: &{\he}3.595 &{\he}3.475 &3.385 &$\hc0.128$ &0.720 &0.177&0.284 \cr
\+{\hf}1988: &22.810  &22.469  &1.670  &$-0.147$   &0.701  &0.335  &0.513 \cr
\+{\hf}1987: &18.461  &17.823  &2.056  &$-0.151$   &0.553  &0.326  &0.362 \cr
\+{\hf}1986: &16.950  &17.009  &2.101  &$\hc0.035$ &0.519  &0.256  &0.266 \cr
\+{\hf}1985: &13.093  &12.803  &2.537  &$\hc0.090$ &0.517  &0.219  &0.227 \cr
\+{\hf}1984: &17.507  &17.136  &2.088  &$\hc0.131$ &0.847  &0.134  &0.314 \cr

\medskip\noindent 
{\csc Table 3.} 
{\sl Estimates of the four mean value parameters $a_1$, $a_2$, $b$, $d$
and of the three variance-covariance parameters $\rho$, $\sigma$, $\kappa$,
for each of the eleven Sprint World Championships for men, 1984--1994.}
\smallskip}}

It is also illuminating to present evidence against 
the $d=0$ hypothesis in terms of direct as well as 
suitably corrected and standardised differences in lap times. 
The lap time difference for skater $i$ is 
$r_{2,i}-r_{1,i}=a_2-a_1+(b-1)(x_{2,i}-x_{1,i})+e_{2,i}-e_{1,i}$,
if $d=0$, so that the variable 
$$\eqalign{D_i
&=r_{2,i}-r_{1,i}-\{\hatt a_2-\hatt a_1+(\hatt b-1)(x_{2,i}-x_{1,i})\} \cr 
&=(Y_{2,i}-\hatt a_2-\hatt b x_{2,i})
       -(Y_{1,i}-\hatt a_1-\hatt b x_{1,i}) \cr} \eqno(5.2)$$
represents adjusted lap time difference for each skater,
where `adjusted' means relative to varying ice and weather conditions 
on the two days and information contained in the 100 meter passing times, 
but {\it not} adjusted for inner-outer lane information. 
The $D_i$s are directly interpretable on the original time scale in seconds. 
Figure 3 gives density estimates for 
$D_i$s observed for two groups of skaters, for each of the SWCs 1984--1994. 
The first group is the one with $(z_{1,i},z_{2,i})=(1,-1)$, 
or $w_i=\half$, and has the presumed preferable last outer lane 
on the second day; 
the complimentary group has $(z_{1,i},z_{2,i})=(-1,1)$, 
or $w_i=-\half$, with last outer lane on the first day. 
Accordingly, {\it if} there is unfairness in the expected direction 
($d$ positive), then the $D_i$s for the $w_i=\half$ group can be 
expected to lean slightly to the left 
(mean value around $-d$), 
while the $D_i$s for the $w_i=-\half$ group would tend to 
lean slightly to the right (mean value around $d$). 
In Figure 3 skaters who fell or were disqualified were 
eliminated, as were those failing one or more of the three outlier 
tests described in Section 4. 
When presenting these pairs of densities I also went 
to the trouble of calculating the $D_i$s with 
recomputed estimates $(\hatt a_1,\hatt a_2,\hatt b)$ 
\fermat{Figure 3 with its eleven parts 
is found on pages 24-25}under the $d=0$ hypothesis.

These figures give good visual impressions of the tentative 
differences between the inner and outer lane situations, 
for each particular SWC 1984--1994. 
They also reveal information about variability level,
corresponding to the $\hatt\sigma_{\rm un}$ values given in Table 3 above;
Troms\o{} 1990, West Allis 1988 and Sainte Foy 1987 were quite
variable occasions (again, explainable by weather and gliding conditions),
whereas Trondheim 1994, Heerenveen indoor 1989 and Inzell 1991,
for example, were `cleaner' occasions with 
less variability around each skater's normal capacity level. 
Figure 4 presents essentially the same information in another way,
by ordinary data dot plots. 

{{\smallskip\narrower\baselineskip11pt\noindent 
{\csc Figure 3.} 
{\sl Nonparametric kernel-method density estimates 
for the modified difference variables $D_i$, 
for the $w_i=1/2$ group (fully drawn) and for the $w_i=-1/2$ group 
(dotted line), for each of the eleven SWCs 1984--1994. 
The advantage of having last outer lane is arguably prominent 
for the 1991, 1990, 1989, 1986, 1985, 1984 occasions, 
undecided for the 1994, 1993, 1992 events, 
while the advantage seems to have been 
with the last inner lane in 1988 and 1987. 
The number of skaters contributing to the eleven sub-figures  
are respectively 26, 29, 30, 33, 28, 29, 28, 32, 26, 30, 27, 
for the years 1984--1994.}
\medskip}}

\medskip 
\centerline{\includegraphics[scale=0.66]{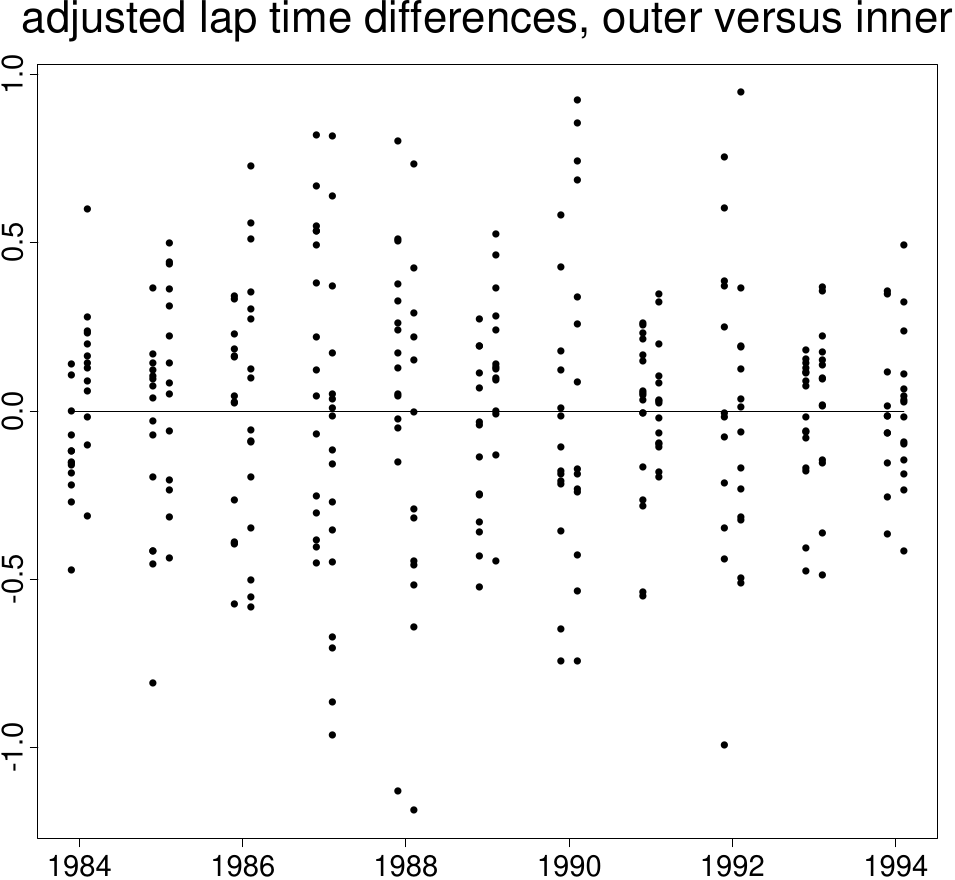}}

{{\medskip\narrower\baselineskip11pt\noindent 
{\csc Figure 4.} 
{\sl Plots of the modified difference variables $D_i$, 
for the $w_i=1/2$ group (to the left) and for the $w_i=-1/2$ group 
(to the right), for each of the eleven SWCs 1984--1994. 
See also the legend to Figure 3.}
\smallskip}}

A quite informative overall picture can also be formed 
by comparing the all in all 159 skaters who had $w_i=\half$ 
with the 159 skaters who had $w_i=-\half$.
The comparison is most meaningful if the $D_i$s are all scaled
with the appropriate estimated standard deviation, 
which varies from year to year. Figure 5 therefore presents 
densities for the two groups, of the adjusted and standardised variables 
$$\eqalign{D_i^*=D_i/\sqrt{2}\hatt\sigma_{\rm un}
&=\bigl[r_{2,i}-r_{1,i}-\{\hatt a_2-\hatt a_1
        +(\hatt b-1)(x_{2,i}-x_{1,i})\}\bigr]
        /\sqrt{2}\hatt\sigma_{\rm un} \cr
&=\{(Y_{2,i}-\hatt a_2-\hatt b x_{2,i})-(Y_{1,i}-\hatt a_1-\hatt b x_{1,i})\}
        /\sqrt{2}\hatt\sigma_{\rm un}, \cr} \eqno(5.3)$$
with the appropriate year-specific $\hatt\sigma_{\rm un}$. 
Again the $D^*_i$s that were used employed re-estimated 
versions of $\hatt a_1$, $\hatt a_2$, $\hatt b$, $\hatt\sigma_{\rm un}$,
all arrived at under the $d=0$ hypothesis. 
The $D_i^*$s with $w_i=\half$ should lean somewhat to the left
(mean value around $-d/\sqrt{2}\sigma$) while those with $w_i=-\half$
should lean the other way (mean value around $d/\sqrt{2}\sigma$),
again, {\it if}\/ $d$ indeed is positive. 
The figure seems to support this, 
in view of the large sample sizes for the two group. 
If my reader agrees that the $w_i=\half$ curve gives 
slightly but markedly more probability mass to 
the left hand side than does the $w_i=-\half$ curve, 
then he or she might endorse the change in rule towards asking 
the skaters to sprint twice. 

\medskip
\centerline{\includegraphics[scale=0.66]{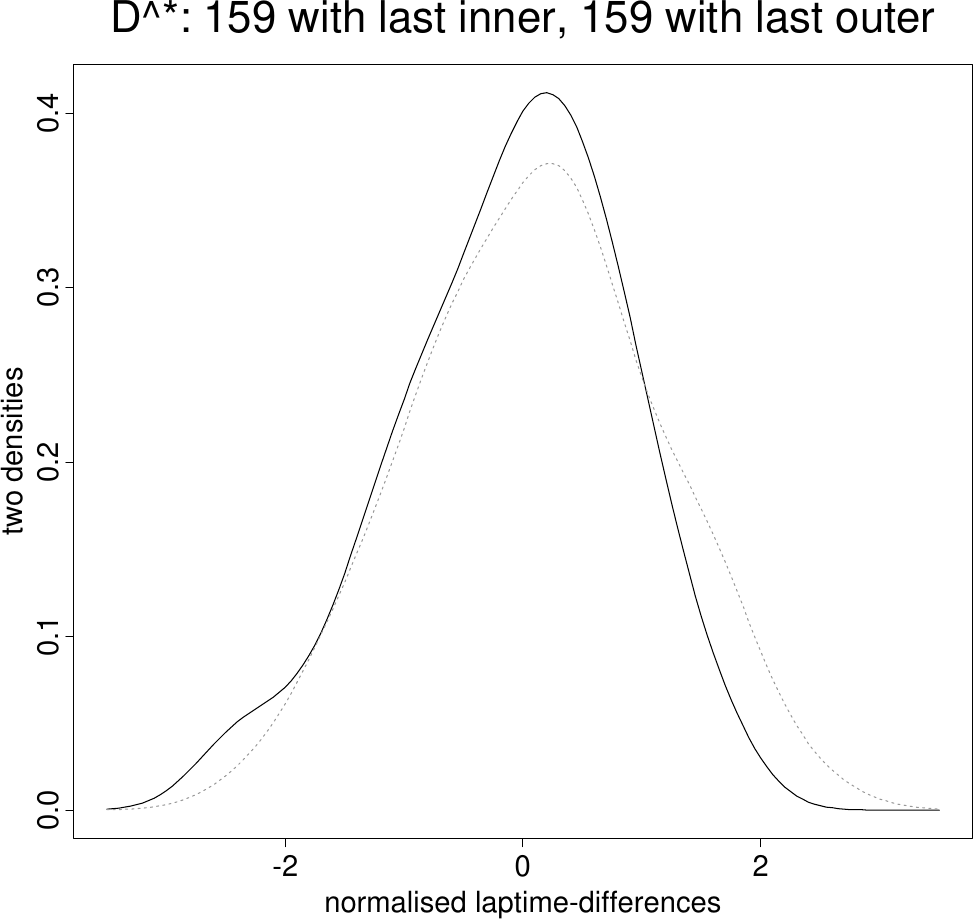}}


{{\medskip\narrower\baselineskip11pt\noindent 
{\csc Figure 5.}
{\sl  
Density estimates for the two groups of adjusted and standardised 
difference variables $D_i^*$ of (5.3). 
The fully drawn curve is that of the 159 skaters 
who had last outer lane at the Sunday event 
while the dotted curve is that of the complimentary 
159 skaters who had last outer lane on Saturday. 
The former gives more probability mass to the left,
indicating that the last outer lane is advantageous.}  
\smallskip}}
\eject 

\centerline{\bf 6. Women} 

\medskip\noindent 
Women are different from men, aerodynamically speaking. 
As hinted at in the introduction section 
the best women sprinters do not seem to be hampered as much as 
many male sprinters with high speed in the last inner lane,
and also offer spectators far fewer spectacular high speed falls
(there were only three 500 meter-falls in eleven SWCs among the women,
including Christa Rothenburger's in 1985, where she still managed 
to win the Championship, but 16 in the same period among the men). 
I nevertheless went ahead to collect and analyse the same amount of 
data on the women's runs, 
adhering to the Equal Statistics for Women commandment, 
to estimate also their seemingly less significant $d$ parameter. 
The results were as follows, as far as the $d$ estimation is concerned. 
Note that the SWCs are always held jointly for men and women, 
on the same days, so the same weather and ice quality conditions 
reign over both. 

\medskip
\centerline{\includegraphics[scale=0.66]{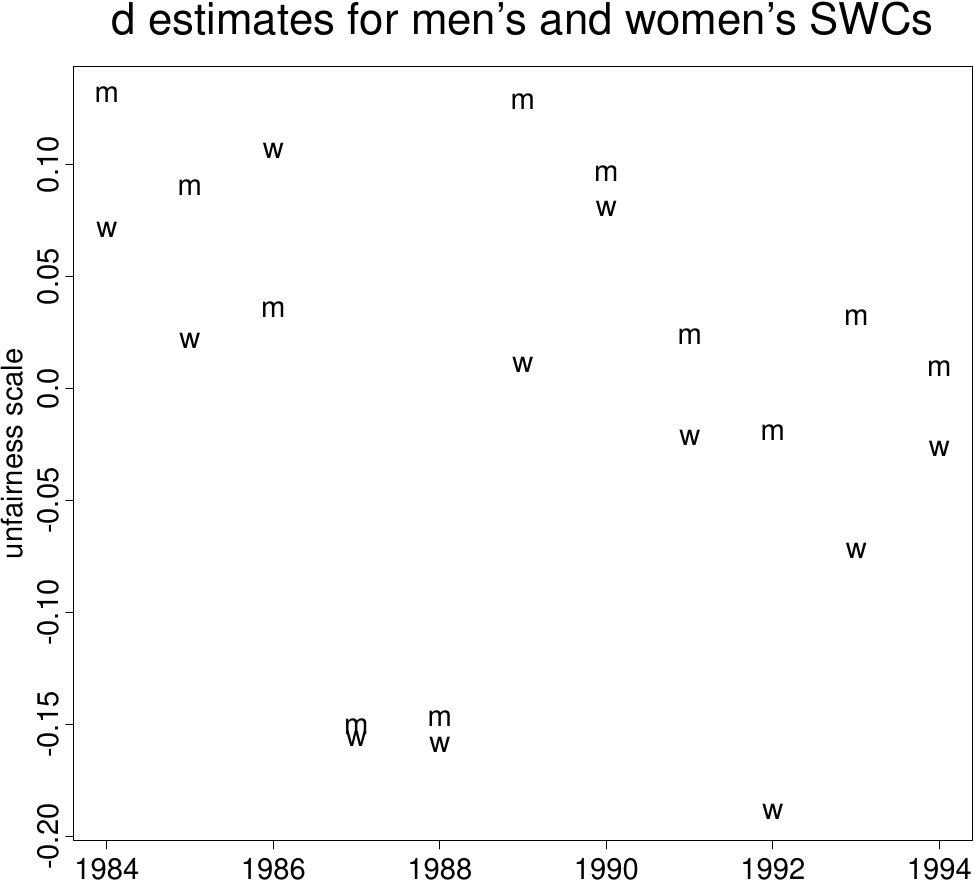}}


{{\medskip\narrower\baselineskip11pt\noindent 
{\csc Figure 6.} {\sl 
Estimates of the $d$ parameter, for the women's and the men's events,
for eleven SWCs 1984--1994. The correlation is 0.792, and indicates
that the real $d$ parameter changes somewhat from event to event.} 
\smallskip}}

The first thing to note is that it is not possible to 
reject the null hypothesis $d=0$ based on these data,
and any reasonable confidence interval will cover zero. 
A second prominent feature is that the $d$ estimates for women
are highly correlated with those for the men 
(the correlation coefficient is 0.792); see Figure 6. 
This supports the notion that each SWC event 
has its own characteristics, depending on wind, 
weather and gliding conditions; in addition two skating
rinks might differ somewhat with respect to the actual radii used
for inner and outer lanes. This also lends support to the 
appropriateness of the new ISU/IOC rules, 
that also the women skaters are to run the Olympic 500 meter twice 
from Nagano 1998 onwards, as well as in the annual single distance 
World Championships that are introduced in 1996. 

{{\smallskip\baselineskip11pt\narrower
\def\hf{\hfill}
\def\ha{\hskip1.50cm} 
\def\hb{\hskip1.25cm} 
\def\hc{\phantom{-}}
\settabs\+\ha\hf 1989 &Heerenveen: \hb&--0.019 \hskip0.50cm &0.086 \cr 
\+\hf      &            &$\hc\hc \hatt d$  &$\!\!$st.~error \cr 

\smallskip
\+\hf 1994 &Calgary:    &$   -0.027$  &0.039 \cr 
\+\hf 1993 &Ikaho:      &$   -0.072$  &0.067 \cr  
\+\hf 1992 &Oslo:       &$   -0.189$  &0.090 \cr   
\+\hf 1991 &Inzell:     &$   -0.022$  &0.079 \cr  
\+\hf 1990 &Troms\o:    &$\hc 0.080$  &0.063 \cr 
\+\hf 1989 &Heerenveen: &$\hc 0.010$  &0.046 \cr 
\+\hf 1988 &West Allis: &$   -0.159$  &0.086 \cr 
\+\hf 1987 &Sainte Foy: &$   -0.157$  &0.159 \cr
\+\hf 1986 &Karuizawa:  &$\hc 0.106$  &0.095 \cr  
\+\hf 1985 &Heerenveen: &$\hc 0.021$  &0.079 \cr
\+\hf 1984 &Trondheim:  &$\hc 0.071$  &0.081 \cr  
\+\hf      &grand average: &$-0.015$  &0.020 \cr 

\smallskip\noindent
{\csc Table 4.} 
{\sl Estimates of the difference parameter $d$ for eleven 
Sprint World Championships for women, along with standard errors 
(estimated standard deviation) for these.}
\smallskip}}


\bigskip
\centerline{\bf 7. Supplementing remarks} 

\medskip
{\csc Remark A.} 
The inner-outer question can also be asked for the 1000 meter 
competition. In this event skaters the first 200 meter, 
in particular, are quite different stretches for the two skaters.
A potential difference is not likely to be statistically visible,
however; the distance is long enough to diminish any tiny discrepancies. 
My student Siri St\o rmer is considering SWC results 
from several 1000 meters, looking for inner-outer significances
as well as other aspects, and indeed preliminary investigations 
have not found any significant inner-outer differences.  
 


\subsection
{\csc Remark B.} 
It is natural to think that the $d$ parameter is not quite 
one and only one constant, but that it changes slightly from event
to event, depending on weather and the rink itself. 
The comments regarding the similarity between the men's
and the women's $d$ estimates above supports this notion, cf.~Figure 6. 
A simple model for this is to postulate that 
$d_j\sim\normal\{d_0,\omega_0^2\}$, where $d_0$ is some presumed 
grand mean over many events and $\omega_0$ the level of variation. 
Our $\hatt d_j$ is approximately a $\normal\{d_j,{\rm se}_j^2\}$,
conditionally on $d_j$. One may now show that 
$$\sum_{j=1984}^{1994}(\hatt d_j-\hatt d)^2/{\rm se}_j^2
        \quad {\rm has\ mean\ value} \quad 
        10+\omega_0^2(A_2-A_4/A_2), $$
where $\hatt d$ is as in (5.1) and 
$A_q=\sum_{j=1984}^{1994}1/{\rm se}_j^q$ for $q=2,4$.  
This gives us the opportunity to 
estimate the variability between events parameter $\omega_0$;
I find 0.057 for the men's events and 0.042 for the women's. 
In other words, according to this model, 
the $d_j$s can be expected to wander from about $-0.05$ up to about $0.14$,
in 90\% of such competitions for the men.
The corresponding figures for the women are $-0.08$ to $0.05$.

That the underlying parameters of the model,
from $(a_1,a_2,b,d)$ to $(\sigma,\rho,\kappa)$,
vary from event to event in a suitably regular fashion 
can be modelled as well, as with the $d_j$ parameters. 
This may lead us to empirical Bayes modelling and estimation.
Such a framework yields refined estimators of the individual 
parameters that in an overall sense would be more precise 
than for example the $\hatt d_j$s of Section 1's Table 1. 
Our main task has been to estimate the grand average of these,
however, and it seems more natural to let the parameters 
of each competition speak for themselves, without weighing in 
similar information from other years. 

\subsection
{\csc Remark C.} 
One may ask whether there is a difference between the very 
best skaters in the world and the not quite as excellent ones,
regarding their ability to tackle the last inner lane. 
The answer would depend on the selected party with which 
one wishes to compare the very best. 
The level of the skaters being allowed to compete 
at the SWCs is now uniformly very high, however, 
and significant differences would have been surprising. 
To investigate this matter, I divided each of the 1984--1994 sets 
into two halves, the best part (decided on by the average 500 meter result) 
and the remaining ones. For each half one can fit the (2.2) model
(with $b_1=b_2)$, and in particular compare the $\hatt d_{\rm best}$ 
estimate for the very best skaters in the world with 
the corresponding $\hatt d_{\rm rest}$ valid for the not quite as 
spectacular skaters. Overall there seems to be a certain tendency 
towards $d_{\rm best}<d_{\rm rest}$, that is, 
the very best skaters are better at handling also the last inner lane 
problems. This discrepancy is not significant, however;
there are instances where the opposite happens, 
and the best combined estimate of $d_{\rm best}-d_{\rm rest}$ is $-0.044$
with standard error $0.029$. 


\bigskip
{\bf Acknowledgements.} 
I am indebted to all the participants of the statistical experiments 
described here for their eager contributions. 
I am also grateful for the interest shown in this project by Tron Espeli 
at the International Skating Union's Technical Committee for Speed Skating. 

\bigskip
\centerline{\bf Reference} 

\parindent0pt
\baselineskip11pt
\parskip3pt 

\medskip 
\ref{Bj\o rnsen, K. (1963).
{\sl 13 \aa r med Kuppern \& Co.}
Nasjonalforlaget, Oslo.
[`Kuppern' is Knut Johannesen, Olympic medalist in 1956, 1960, 1964.]} 




\vfill\eject 

\centerline{\bf Appendix I} 

\medskip\noindent 
The left hand list given here gives the real results from the Olympic
500 meter sprint event. The supplementary faked list is meant 
to show what presumably would have happened if the 
inner-outer lane draw had ended in the opposite way. 
For example, 1988 silver medalist Jan Ykema would have ended fourth
while fourth ranked Sergei Fokitshev would have grabbed the bronze, 
and so on. 

\def\h{\hskip0.33cm} 
\def\hb{\hskip0.22cm} 
\def\hf{\hfill}  
\def\in{\hskip0.04truecm i}
\baselineskip11pt 

\settabs\+\hf 20. &Hanspeter Oberhuber \h &i \hb &37.73 \qquad 
                &\hf 20. &Hanspeter Oberhuber \h &37.79 \cr 
\bigskip
\+\hf &{\bf Olympic Games, Lillehammer 1994} \cr
\+\hf &{\bf Real list:} && 
                & &{\bf Speculative list:} \cr 

\smallskip  
\+\hf 1. &Aleksandr Golubyev   &\in &36.33 
                &\hf 1. &Aleksandr Golubyev     &36.38 \cr 
\+\hf 2. &Sergei Klevtshena      &\in &36.39 
                &\hf 2. &Sergei Klevtshena      &36.44 \cr
\+\hf 3. &Manabu Horii           &o   &36.53 
                &\hf 3. &Manabu Horii           &36.48 \cr
\+\hf 4. &Hongbo Liu             &\in &36.54 
                &\hf 4. &Hiroyasu Shimizu       &36.55 \cr
\+\hf 5. &Hiroyasu Shimizu       &o   &36.60 
                &\hf 5. &Hongbo Liu             &36.59 \cr
\+\hf 6. &Junichi Inoue          &\in &36.63 
                &\hf 6. &Grunde Nj\o s          &36.61 \cr
\+\hf 7. &Grunde Nj\o s          &o   &36.66 
                &\hf 7. &Yasunori Miyabe        &36.67 \cr
\+\hf 8. &Dan Jansen             &\in &36.68 
                &\hf 8. &Junichi Inoue          &36.68 \cr
\+\hf 9. &Yasunori Miyabe        &o   &36.72 
                &\hf    &Igor Zhelezovsky       &36.68 \cr
\+\hf 10. &Igor Zhelezovsky      &o   &36.73 
                &\hf 10. &Dan Jansen            &36.73 \cr
\+\hf 11. &Sylvain Bouchard      &o   &37.01 
                &\hf 11. &Sylvain Bouchard      &36.96 \cr
\+\hf 12. &Patrick Kelly         &\in &37.07 
                &\hf 12. &Yoon-Man Kim          &37.05 \cr 
\+\hf     &Vadim Shakshakbayev   &\in &37.07 
                &\hf 13. &Patrick Kelly         &37.12 \cr
\+\hf 14. &Yoon-Man Kim          &o   &37.10 
                &\hf     &Vadim Shakshakbayev   &37.12 \cr
\+\hf 15. &Mikhail Vostroknutov  &\in &37.15 
                &\hf 15. &Mikhail Vostroknutov  &37.20 \cr
\+\hf 16. &Andrei Bakhvalov      &\in &37.24 
                &\hf 16. &Sean Ireland          &37.25 \cr
\+\hf 17. &Sean Ireland          &o   &37.30 
                &\hf 17. &Andrei Bakhvalov      &37.29 \cr
\+\hf 18. &Peter Adeberg         &o   &37.35 
                &\hf 18. &Peter Adeberg         &37.30 \cr
\+\hf 19. &David Cruikshank      &\in &37.37 
                &\hf 19. &Nathaniel Mills       &37.36 \cr
\+\hf 20. &Nathaniel Mills       &o   &37.41 
                &\hf 20. &Gerard van Velde      &37.40 \cr
\+\hf 21. &Gerard van Velde      &o   &37.45 
                &\hf 21. &David Cruikshank      &37.42 \cr
\+\hf 22. &Roland Brunner        &\in &37.47 
                &\hf 22. &Arie Loef             &37.47 \cr
\+\hf 23. &Oleg Kostromitin      &\in &37.50 
                &\hf 23. &Hans Markstr\"om      &37.48 \cr
\+\hf 24. &Arie Loef             &o   &37.52 
                &\hf 24. &Roland Brunner        &37.52 \cr
\+\hf 25. &Hans Markstr\"om      &o   &37.53 
                &\hf 25. &Oleg Kostromitin      &37.55 \cr
\+\hf 26. &Michael Ireland       &\in &37.67 
                &\hf 26. &Michael Ireland       &37.72 \cr
\+\hf 27. &David Besteman        &\in &37.68 
                &\hf 27. &David Besteman        &37.73 \cr
\+\hf 28. &Lars Funke            &o   &37.80 
                &\hf 28. &Lars Funke            &37.75 \cr
\+\hf 29. &Alessandro de Taddei  &\in &37.87 
                &\hf 29. &Sung-Yeol Jaegal      &37.85 \cr
\+\hf 30. &Sung-Yeol Jaegal      &o   &37.90 
                &\hf 30. &Nico van der Vlies    &37.89 \cr
\+\hf 31. &Nico van der Vlies    &o   &37.94 
                &\hf 31. &Alessandro de Taddei  &37.92 \cr
\+\hf 32. &Davide Carta          &o   &37.98 
                &\hf 32. &Davide Carta          &37.93 \cr
\+\hf 33. &Vladimir Klepinin     &o   &38.09 
                &\hf 33. &Vladimir Klepinin     &38.02 \cr
\+\hf 34. &Magnus Enfeldt        &\in &38.10 
                &\hf 34. &Jae-Shik Lee          &38.05 \cr
\+\hf     &Jae-Shik Lee          &o   &38.10 
                &\hf 35. &Kyou-Hyuk Lee         &38.08 \cr
\+\hf 36. &Kyou-Hyuk Lee         &o   &38.13 
                &\hf 36. &Magnus Enfeldt        &38.15 \cr
\+\hf 37. &Arjan Schreuder       &\in &38.33 
                &\hf 37. &Arjan Schreuder       &38.38 \cr
\+\hf 38. &Zsolt Balo            &\in &38.56 
                &\hf 38. &Zsolt Balo            &38.61 \cr
\+\hf 39. &Michael Spielmann     &\in &38.58 
                &\hf 39. &Michael Spielmann     &38.63 \cr
\+\hf --\h &Roger Str\o m         &\in &dnf 
                &\hf     &Roger Str\o m & --- \cr 

\def\h{\hskip0.33cm} 
\def\hb{\hskip0.22cm} 
\def\hf{\hfill}  
\def\in{\hskip0.04truecm i}
\baselineskip11pt 

\settabs\+\hf 20. &Hanspeter Oberhuber \h &i \hb &37.73 \qquad 
                &\hf 20. &Hanspeter Oberhuber \h &37.79 \cr 

\bigskip
\+\hf &{\bf Olympic Games, Albertville 1992} \cr
\+\hf &{\bf Real list:} && 
                & &{\bf Speculative list:} \cr

\smallskip
\+\hf  1. &Uwe-Jens Mey         &o &37.14 
                &\hf 1.  &Uwe-Jens Mey          &37.09 \cr
\+\hf  2. &Toshiyuki Kuroiwa    &\in &37.18 
                &\hf 2.  &Toshiyuki Kuroiwa     &37.23 \cr                     
\+\hf  3. &Junichi Inoue        &\in &37.26     
                &\hf 3.  &Junichi Inoue         &37.31 \cr
\+\hf  4. &Dan Jansen           &o &37.46
                &\hf 4.  &Dan Jansen            &37.41 \cr  
\+\hf  5. &Yasunori Miyabe      &\in &37.49 
                &\hf 5.  &Gerard van Velde      &37.44 \cr
\+\hf     &Gerard van Velde     &o &37.49
                &\hf 6.  &Aleksandr Golubyev    &37.46 \cr
\+\hf  7. &Aleksandr Golubyev   &o &37.51
                &\hf 7.  &Chen Song             &37.53 \cr
\+\hf  8. &Igor Zhelezovsky     &\in &37.57
                &\hf 8.  &Yasunori Miyabe       &37.54 \cr 
\+\hf  9. &Chen Song            &o &37.58
                &\hf 9.  &Yoon-Man Kim          &37.55 \cr
\+\hf 10. &Yoon-Man Kim         &o &37.60       
                &\hf 10. &Igor Zhelezovsky      &37.62 \cr
\+\hf 11. &Hongbo Liu           &\in &37.66
                &\hf 11. &Sung-Yul Yegal        &37.66 \cr
\+\hf 12. &Sung-Yul Yegal       &o &37.71
                &\hf 12. &Hongbo Liu            &37.71 \cr 
\+\hf 13. &Nick Thometz         &\in &37.83
                &\hf 13. &Vadim Shakshakbayev   &37.81 \cr
\+\hf 14. &Robert Dubreuil      &\in &37.86
                &\hf 14. &Guy Thibault          &37.84 \cr 
\+\hf     &Vadim Shakshakbayev  &o &37.86 
                &\hf 15. &Nick Thometz          &37.88 \cr 
\+\hf 16. &Guy Thibault         &o &37.89
                &\hf 16. &Robert Dubreuil       &37.91 \cr 
\+\hf 17. &Kevin Scott          &\in &38.02
                &\hf 17. &Kevin Scott           &38.07 \cr 
\+\hf 18. &Yukihiro Miyabe      &\in &38.12
                &\hf 18. &Yukihiro Miyabe       &38.17 \cr 
\+\hf 19. &Marty Pierce         &\in &38.15
                &\hf 19. &Marty Pierce          &38.20 \cr 
\+\hf 20. &Bj\"orn Forslund     &\in &38.24
                &\hf 20. &Peter Adeberg         &38.28 \cr  
\+\hf 21. &Sergei Klevtshena   &\in &38.26
                &\hf 21. &Bj\"orn Forslund      &38.29 \cr 
\+\hf 22. &David Cruikshank     &\in &38.28
                &\hf 22. &Sergei Klevtshena     &38.31 \cr
\+\hf 23. &Peter Adeberg        &o &38.33
                &\hf 23. &Yong-Cho Li           &38.33 \cr 
\+\hf 24. &Yong-Cho Li          &o &38.38
                &\hf     &David Cruikshank      &38.33 \cr
\+\hf 25. &Olaf Zinke           &o &38.40
                &\hf 25. &Olaf Zinke            &38.35 \cr
\+\hf 26. &Harri Ilkka          &\in &38.48
                &\hf 26. &Rintje Ritsma         &38.46 \cr 
\+\hf 27. &Jun Dai              &\in &38.51
                &\hf 27. &Harri Ilkka           &38.53 \cr  
\+\hf     &Rintje Ritsma        &o &38.51
                &\hf 28. &Arie Loef             &38.56 \cr 
\+\hf 29. &Arie Loef            &o &38.61
                &\hf     &Jun Dai               &38.56 \cr  
\+\hf 30. &Sean Ireland         &\in &38.70
                &\hf 30. &Pawel Abratkiewicz    &38.69 \cr 
\+\hf 31. &Pawel Abratkiewicz   &o &38.74
                &\hf     &In-Hoon Lee           &38.69 \cr 
\+\hf     &In-Hoon Lee          &o &38.74
                &\hf 32. &Sean Ireland          &38.75 \cr 
\+\hf 33. &Hans Markstr\"om     &o &38.89
                &\hf 33. &Hans Markstr\"om      &38.84 \cr 
\+\hf 34. &Bo K\"onig           &\in &39.06
                &\hf 34. &Bo K\"onig            &39.11 \cr 
\+\hf 35. &In-Hol Choi          &o &39.59
                &\hf 35. &In-Hol Choi           &39.54 \cr
\+\hf 36. &Zsolt Ballo          &\in &39.70 
                &\hf 36. &Zsolt Ballo           &39.75 \cr
\+\hf 37. &Csaba Madarasz       &\in &40.41
                &\hf 37. &Csaba Madarasz        &40.46 \cr
\+\hf 38. &Joakim Karlberg      &o &40.71
                &\hf 38. &Joakim Karlberg       &40.66 \cr 
\+\hf 39. &Jiri Kyncl           &\in &40.92
                &\hf 39. &Jiri Kyncl            &40.97 \cr 
\+\hf 40. &Roland Brunner       &o &42.18 
                &\hf 40. &Roland Brunner        &42.13 \cr 
\+\hf 41. &Jiri Musil           &o &42.20 
                &\hf 41. &Jiri Musil            &42.15 \cr 
\+\hf 42. &Bajro Cenanovic      &\in &43.09 
                &\hf 42. &Bajro Cenanovic       &43.14 \cr 
\+\hf 43. &Slavenko Likic       &o &43.81 
                &\hf 43. &Slavenko Likic        &43.76 \cr 

\def\h{\hskip0.33cm} 
\def\hb{\hskip0.22cm} 
\def\hf{\hfill}  
\def\in{\hskip0.04truecm i}
\baselineskip11pt 

\settabs\+\hf 20. &Hanspeter Oberhuber \h &i \hb &37.73 \qquad 
                &\hf 20. &Hanspeter Oberhuber \h &37.79 \cr 

\bigskip
\+\hf &{\bf Olympic Games, Calgary 1988} \cr
\+\hf &{\bf Real list:} && 
                & &{\bf Speculative list:} \cr

\smallskip
\+\hf  1. &Uwe-Jens Mey         &\in &36.45 
                &\hf 1.  &Uwe-Jens Mey          &36.50 \cr
\+\hf  2. &Jan Ykema            &\in &36.76
                &\hf 2.  &Akira Kuroiwa         &36.72 \cr                     
\+\hf  3. &Akira Kuroiwa        &o &36.77       
                &\hf 3.  &Sergei Fokitchev      &36.77 \cr              
\+\hf  4. &Sergei Fokitchev     &o &36.82
                &\hf 4.  &Jan Ykema             &36.81 \cr              
\+\hf  5. &Ki-Tae Bae           &o &36.90
                &\hf 5.  &Ki-Tae Bae            &36.85 \cr 
\+\hf  6. &Igor Zhelezovsky     &o &36.94
                &\hf 6.  &Igor Zhelezovsky      &36.89 \cr 
\+\hf  7. &Guy Thibault         &o &36.96
                &\hf 7.  &Guy Thibault          &36.91 \cr 
\+\hf  8. &Nick Thometz         &\in &37.16
                &\hf 8.  &Nick Thometz          &37.21 \cr 
\+\hf  9. &Yasumitsu Kanehama   &\in &37.25
                &\hf 9.  &Yasushi Kuroiwa       &37.29 \cr  
\+\hf 10. &Frode R\o nning      &\in &37.31     
                &\hf 10.  &Vitali Makovetski   &37.30 \cr 
\+\hf 11. &Yasushi Kuroiwa      &o &37.34
                &\hf      &Yasumitsu Kanehama   &37.30 \cr 
\+\hf 12. &Vitali Makovetski   &o &37.35
                &\hf 12.  &Kimihiro Hamaya      &37.33 \cr 
\+\hf 13. &Kimihiro Hamaya      &o &37.38
                &\hf 13.  &Frode R\o nning      &37.36 \cr 
\+\hf 14. &Gaetan Boucher       &\in &37.47 
                &\hf 14.  &Menno Boelsma        &37.47 \cr 
\+\hf 15. &Erik Henriksen       &\in &37.50
                &\hf 15.  &Gaetan Boucher       &37.52 \cr 
\+\hf 16. &Menno Boelsma        &o &37.52
                &\hf 16.  &Erik Henriksen       &37.55 \cr 
\+\hf 17. &Daniel Turcotte      &\in &37.60
                &\hf 17.  &Bj\o rn Hagen        &37.64 \cr 
\+\hf 18. &G\"oran Johansson    &\in &37.69
                &\hf 18.  &Daniel Turcotte      &37.65 \cr 
\+\hf     &Bj\o rn Hagen        &o &37.69
                &\hf 19.  &Marty Pierce         &37.71 \cr 
\+\hf 20. &Hanspeter Oberhuber  &\in &37.73
                &\hf 20.  &Michael Richmond     &37.72 \cr 
\+\hf 21. &Andr\'e Hoffmann     &\in &37.75
                &\hf 21.  &G\"oran Johansson    &37.74 \cr 
\+\hf 22. &Marty Pierce         &o &37.76
                &\hf 22.  &Hein Vergeer         &37.75 \cr 
\+\hf 23. &Michael Richmond     &o &37.77
                &\hf 23.  &Jerzy Dominik        &37.78 \cr 
\+\hf 24. &Hein Vergeer         &o &37.80
                &\hf      &Hanspeter Oberhuber   &37.78 \cr 
\+\hf 25. &Jerzy Dominik        &o &37.83
                &\hf 25.  &Andr\'e Hoffmann     &37.80 \cr 
\+\hf 26. &Michael Hadschieff   &\in &37.90
                &\hf 26.  &Michael Hadschieff   &37.95 \cr 
\+\hf 27. &Uwe Streb            &\in &38.03
                &\hf 27.  &Uwe Streb            &38.08 \cr 
\+\hf 28. &Peter Adeberg        &\in &38.11
                &\hf 28.  &Peter Adeberg        &38.16 \cr 
\+\hf 29. &Robert Tremblay      &\in &38.34
                &\hf 29.  &Robert Tremblay      &38.39 \cr 
\+\hf 30. &Hans Magnusson       &o &38.60 
                &\hf 30.  &Hans Magnusson       &38.55 \cr 
\+\hf 31. &Claude Nicoleau      &\in &38.63
                &\hf 31.  &Claes Bengtsson      &38.61 \cr 
\+\hf 32. &Claes Bengtsson      &o &38.66
                &\hf 32.  &Claude Nicoleau      &38.68 \cr 
\+\hf 33. &Hans van Helden      &\in &39.05
                &\hf 33.  &Hans van Helden      &39.10 \cr 
\+\hf 34. &Christian Eminger    &o &39.70
                &\hf 34.  &Christian Eminger    &39.65 \cr 
\+\hf 35. &Behudin Merdovic     &o &fell
                &\hf --\h &Behudin Merdovic     & --- \cr 
\+\hf 36. &Nikolai Gulyayev     &\in &fell 
                &\hf --\h &Nikolai Gulyayev     & --- \cr 
\+\hf --\h &Dan Jansen          &\in &dnf 
                &\hf --\h &Dan Jansen           & --- \cr 

\vfill\eject

\centerline{\includegraphics[scale=0.66]{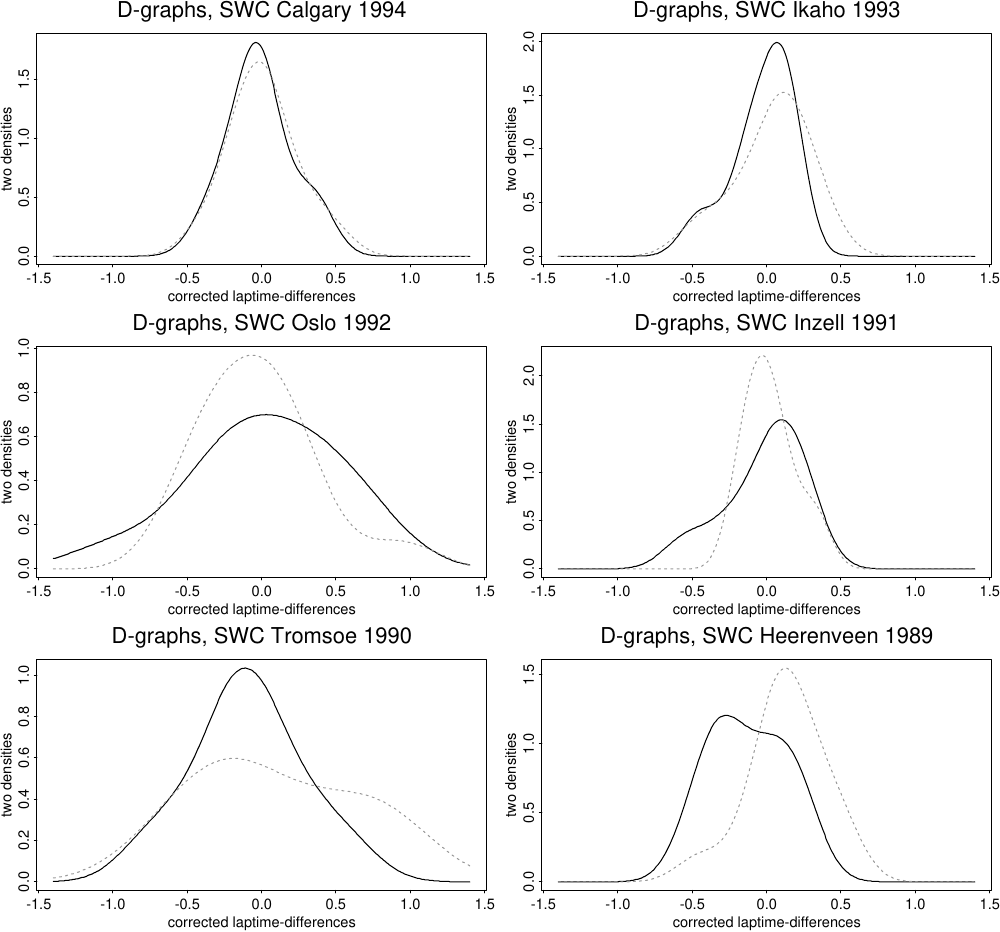}}

\bigskip
\bigskip
{{\smallskip\narrower\baselineskip11pt\noindent 
{\csc Figure 3.} 
{\sl Nonparametric kernel-method density estimates 
for the modified difference variables $D_i$, 
for the $w_i=1/2$ group (fully drawn) and for the $w_i=-1/2$ group 
(dotted line), for each of the eleven SWCs 1984--1994
(six on this page, five on the next). 
The advantage of having last outer lane is arguably prominent 
for the 1991, 1990, 1989, 1986, 1985, 1984 occasions, 
undecided for the 1994, 1993, 1992 events, 
while the advantage seems to have been 
with the last inner lane in 1988 and 1987. 
The number of skaters contributing to the eleven sub-figures  
are respectively 26, 29, 30, 33, 28, 29, 28, 32, 26, 30, 27, 
for the years 1984--1994.}
\smallskip}}

\vfill\eject 

\centerline{\includegraphics[scale=0.66]{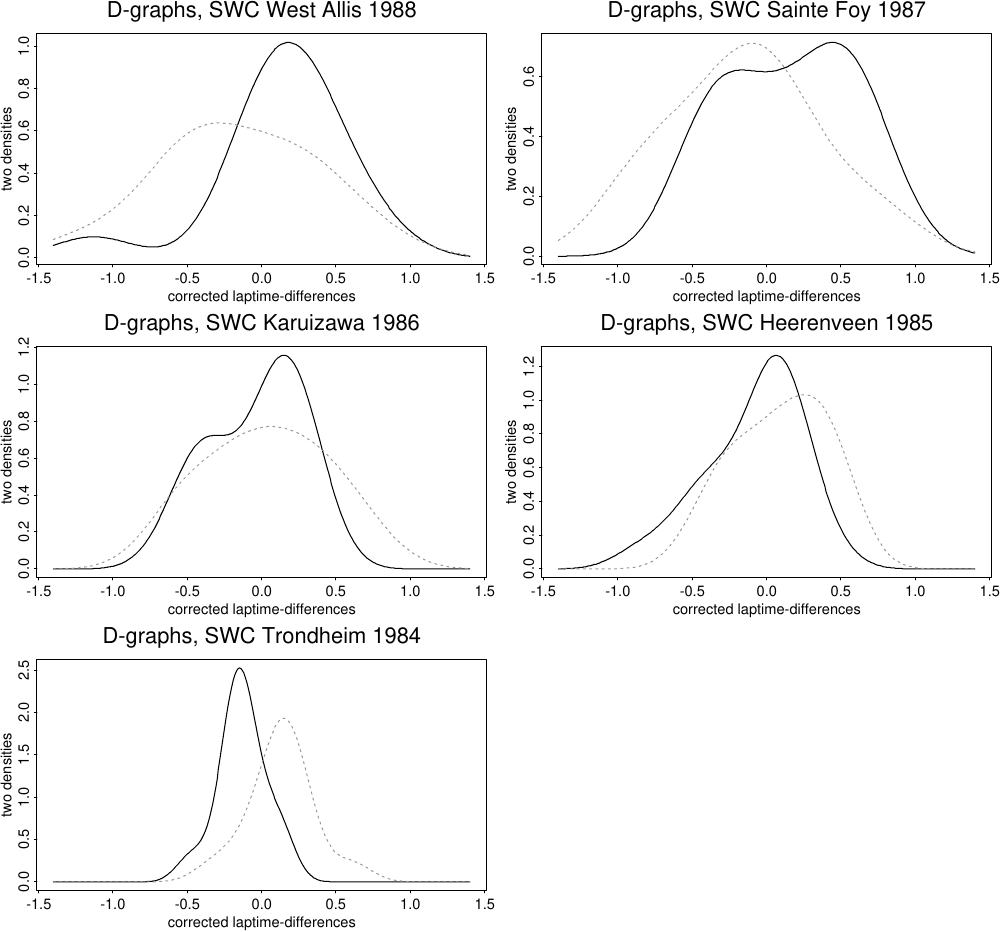}}






\newdimen\fullsize      
\fullsize=16.60truecm \hsize=8.00truecm 
\font\bigbf=cmbx12 
\baselineskip13pt 

\def\fullline{\hbox to \fullsize}
\let\lr=L \newbox\leftcolumn
\output={\if L\lr
                \global\setbox\leftcolumn=\columnbox \global\let\lr=R
        \else \doubleformat \global\let\lr=L\fi
        \ifnum\outputpenalty>-20000 \else\dosupereject\fi}
        \def\doubleformat{\shipout\vbox{\makeheadline
                \fullline{\box\leftcolumn\hfil\columnbox}
                \makefootline}
        \advancepageno}
        \def\columnbox{\leftline{\pagebody}}

\footline={\hskip8.15cm{\rm\folio\rm}\hss}
\supereject
\if R\lr \null\vfill\eject \fi

\def\hf{\hfill}
\def\ha{\hskip0.22cm} 
\def\hb{\phantom{1}} 
\def\in{\hskip0.04cm i}

\font\smallerrm=cmr8

\noindent {\bigbf Appendix II:}

\noindent {\bigbf Complete lists 1984--1994}

\medskip\noindent 
Below the complete results from the 500 meter events of the eleven 
Sprint World Championships 1984--1994 are faithfully recorded.
An `i' signifies starting in inner lane (hence having last outer lane) 
while `o' means starting in outer lane (having last inner lane).
Skaters that were disqualified or fell were removed from the final 
analysis, as were the relatively few who were declared outliers
according to the criteria contructed in Sec\-tion~4. 
These were as follows: 

\smallskip\noindent 
1994: 13 Hamamichi, 27 Konincx and 33 Tahmindjis;  
1992: 25 Dolp;  
1991: 3 T.~Ku\-roiwa and 32 Dolp; 
1989: 24 Sighel; 
1986: 13 Boucher; 
1985: 23 Bae and 31 Henriksen; 
1984: 25 Kuipers. 
There were no additional outliers in 1993, 1990, 1988, 1987. 


\smallerrm
\baselineskip10pt 

\settabs\+\hf 24. &Shakshakbaiev \hb
        &o\ha &10.16\ha &38.49\quad &o\ha &10.02\ha &37.91 \cr 

\medskip
\+\hf     & 
    & &{\smallbf first day:} & & &{\smallbf second day:} \cr 
\+\hf     & 
    & &{\smallbf 100} &{\smallbf 500} & &{\smallbf 100} &{\smallbf 500} \cr 

\medskip
\+ &{\bf Calgary 1994, Jan 29--30:} \cr 

\medskip
\+\hf 1. &D.~Jansen             
        &o &{\hb}9.82 &35.96    &\in &{\hb}9.75 &35.76 \cr 
\+\hf 2. &S.~Klevtshena          
        &o &{\hb}9.78 &36.39    &\in &{\hb}9.82 &36.27 \cr
\+\hf 3. &J.~Inoue              
        &\in &{\hb}9.98 &36.43  &o &{\hb}9.76 &36.05 \cr
\+\hf 4. &H.~Shimizu            
        &o &{\hb}9.70 &36.35    &\in &{\hb}9.77 &36.08 \cr
\+\hf 5. &K.~Scott              
        &\in &{\hb}9.96 &36.87  &o &{\hb}9.87 &36.55 \cr
\+\hf 6. &I.~Zhelezovsky        
        &\in &10.08 &36.90      &o &10.22 &36.99 \cr
\+\hf 7. &T.~Kuroiwa            
        &\in &10.11 &36.83      &o &10.07 &36.75 \cr
\+\hf 8. &Y.-M.~Kim             
        &o &{\hb}9.81 &36.56    &\in &{\hb}9.76 &36.53 \cr
\+\hf 9. &H.B.~Liu              
        &\in &{\hb}9.94 &36.62  &o &10.05 &36.83 \cr
\+\hf 10. &P.~Kelly             
        &o &{\hb}9.97 &37.05    &\in &{\hb}9.92 &36.84 \cr
\+\hf 11. &S.~Bouchard          
        &\in &10.12 &36.97      &o &10.01 &36.73 \cr
\+\hf 12. &G.~Nj\o s                    
        &o &{\hb}9.89 &36.57    &\in &{\hb}9.86 &36.82 \cr
\+\hf 13. &T.~Hamamichi         
        &\in &10.93f &37.50     &o &{\hb}9.69 &36.12 \cr
\+\hf 14. &A.~Golubev           
        &o &{\hb}9.82 &36.60    &\in &{\hb}9.75 &36.76 \cr
\+\hf 15. &Vostroknutov      
        &o &{\hb}9.87 &37.16    &\in &{\hb}9.91 &36.94 \cr
\+\hf 16. &R.~Str\o m           
        &\in &{\hb}9.99 &37.07  &o &10.03 &36.87 \cr
\+\hf 17. &L.~Funke             
        &\in &10.11 &37.35      &o &10.28 &37.23 \cr
\+\hf 18. &R.~Brunner           
        &\in &{\hb}9.94 &37.21  &o &{\hb}9.99 &37.33 \cr
\+\hf 19. &N.~Mills             
        &o &10.12 &37.18        &\in &10.15 &37.18 \cr
\+\hf 20. &A.~Loef              
        &\in &{\hb}9.95 &37.18  &o &{\hb}9.97 &37.50 \cr
\+\hf 21. &A.~Bakhvalov         
        &\in &{\hb}9.96 &37.26  &o &10.14 &37.48 \cr
\+\hf 22. &Shakshakbayev     
        &\in &{\hb}9.91 &36.98  &o &{\hb}9.95 &37.51 \cr
\+\hf 23. &N.~vd Vlies          
        &o &10.25 &37.97        &\in &10.33 &38.07 \cr
\+\hf 24. &W.~Gjevik            
        &\in &10.17 &37.88      &o &10.32 &38.12 \cr
\+\hf 25. &V.~Klepinin          
        &o &10.01 &37.75        &\in &10.06 &37.74 \cr
\+\hf 26. &H.~Ilkka             
        &\in &10.34 &38.05      &o &10.24 &38.04 \cr
\+\hf 27. &P.H.~Koninckx        
        &o &10.17 &38.52        &\in &10.21 &38.58 \cr
\+\hf 28. &H.~Haan              
        &o &10.38 &38.36        &\in &10.35 &38.27 \cr
\+\hf 29. &O.~Kostromitin       
        &o &{\hb}9.90 &37.01    &\in &{\hb}9.93 &fell \cr
\+\hf 30. &G.~v Velde           
        &\in &10.08 &37.04      &o &10.23 &37.11 \cr
\+\hf 31. &H.~Markstr\"om      
        &o &10.06 &fell         &\in &10.19 &37.85 \cr
\+\hf 32. &P.~Hinni             
        &o &10.34 &38.88        &\in &10.26 &fell \cr
\+\hf 33. &P.~Tahmindjis        
        &\in &10.74 &42.47      &o &10.89 &39.09 \cr

\bigskip
\+ &{\bf Ikaho 1993, Feb 26--27:} \cr 

\medskip
\+\hf 1. &I.~Zhelezovsky         
        &\in &10.03 &37.21      &o &10.19 &37.78 \cr 
\+\hf 2. &Yas.~Miyabe            
        &o &{\hb}9.80 &37.09    &\in &{\hb}9.82 &37.18 \cr
\+\hf 3. &H.~Shimizu             
        &\in &{\hb}9.89 &37.13  &o &{\hb}9.77 &37.19 \cr
\+\hf 4. &A.~Golubyev            
        &\in &{\hb}9.93 &37.44  &o &{\hb}9.84 &37.58 \cr
\+\hf 5. &D.~Jansen              
        &o &{\hb}9.92 &37.17    &\in &{\hb}9.98 &37.49 \cr
\+\hf 6. &Yuk.~Miyabe            
        &o &10.04 &37.30        &o &10.00 &37.54 \cr
\+\hf 7. &S.~Ireland             
        &o &10.06 &37.73        &\in &10.10 &37.77 \cr
\+\hf 8. &C.~Tshupira             
        &\in &10.14 &37.80      &o &10.14 &37.78 \cr
\+\hf 9. &H.~Liu                 
        &o &10.06 &37.88        &\in &10.04 &38.13 \cr
\+\hf 10. &S.~Klevtshena          
        &o &10.14 &37.74        &\in &10.11 &37.80 \cr
\+\hf 11. &M.~Horii              
        &\in &10.08 &37.88      &o &10.04 &37.45 \cr
\+\hf 12. &G.~v Velde            
        &\in &10.28 &38.09      &o &10.09 &38.09 \cr
\+\hf 13. &Vostroknutov       
        &o &10.18 &38.38        &\in &{\hb}9.98 &38.08 \cr
\+\hf 14. &R.~Str\o m            
        &o &{\hb}9.97 &37.73    &\in &10.10 &37.70 \cr
\+\hf 15. &G.~Nj\o s              
        &o &10.13 &38.01        &\in & 9.93 &37.59 \cr
\+\hf 16. &P.~Kelly              
        &o &10.25 &38.27        &\in &10.17 &38.38 \cr
\+\hf 17. &Shakshakbayev      
        &o &10.01 &38.04        &\in &10.08 &38.42 \cr
\+\hf 18. &S.-Y.~Jaegal          
        &\in &{\hb}9.90 &37.53  &o &{\hb}9.96 &37.87 \cr
\+\hf 19. &R.~Brunner            
        &o &{\hb}9.93 &37.92    &\in &10.11 &38.33 \cr
\+\hf 20. &A.~Schreuder          
        &\in &10.09 &38.49      &o &10.18 &38.81 \cr
\+\hf 21. &O.~Kostromitin        
        &o &10.02 &38.22        &\in &10.16 &38.14 \cr
\+\hf 22. &M.~Enfeldt           
        &\in &10.42 &38.43      &o &10.25 &38.26 \cr
\+\hf 23. &D.~Besteman           
        &o &10.43 &38.48        &\in &10.36 &38.57 \cr
\+\hf 24. &H.~Markstr\"om       
        &\in &10.31 &38.57      &o &10.47 &38.64 \cr
\+\hf 25. &U.~Streb              
        &\in &10.37 &38.77      &o &10.14 &38.84 \cr
\+\hf 26. &L.~Funke              
        &o &10.14 &38.37        &\in &10.19 &38.71 \cr
\+\hf 27. &M.~Pfeiffer           
        &\in &10.33 &38.71      &o &10.26 &38.81 \cr
\+\hf 28. &P.~Tahmindjis         
        &\in &10.77 &39.71      &o &10.92 &39.98 \cr
\+\hf 29. &D.~Cruikshank         
        &o &10.18 &38.77        &\in &10.15 &38.99 \cr
\+\hf 30. &P.~Seltsam            
        &\in &10.23 &fell       &o &10.29 &38.58 \cr
\+\hf 31. &J.~Wichmann           
        &\in &10.16 &38.84      &didn't start \cr
\+\hf 32. &C.~Song               
        &\in &10.20 &38.01      &o &10.05 &38.22 \cr 

\bigskip
\+ &{\bf Oslo 1992, Feb 29--March 1:} \cr 

\medskip
\+\hf 1. &I.~Zhelezovsky 
        &\in &{\hb}9.89 &37.46  &o   &10.09     &37.72 \cr 
\+\hf 2. &D.~Jansen              
        &\in &{\hb}9.92 &37.92  &o   &10.12     &37.92 \cr
\+\hf 3. &T.~Kuroiwa     
        &\in &10.10     &37.99  &o   &10.12     &37.99 \cr
\+\hf 4. &Yuk.~Miyabe    
        &\in &10.10     &38.55  &o   &10.14     &37.98 \cr
\+\hf 5. &Yas.~Miyabe    
        &o   &{\hb}9.79 &38.60  &\in &{\hb}9.97 &37.88 \cr
\+\hf 6. &G.~v Velde   
        &\in &{\hb}9.98 &38.44  &o   &{\hb}9.99 &38.06 \cr
\+\hf 7. &N.~Gulyayev    
        &o   &10.13     &38.29  &\in &10.03     &38.47 \cr
\+\hf 8. &Shakshakbayev       
        &o   &10.07     &38.37  &\in &10.03     &38.09 \cr
\+\hf 9. &C.~Song                
        &\in &{\hb}9.98 &38.21  &o   &10.03     &38.35 \cr
\+\hf 10. &N.~Thometz            
        &o   &10.07     &38.06  &\in &10.14     &38.43 \cr
\+\hf 11. &O.~Zinke              
        &o   &10.24     &38.63  &\in &10.25     &38.73 \cr
\+\hf 12. &\AA.~S\o ndr\aa l        
        &o   &10.37     &39.50  &\in &10.25     &38.95 \cr
\+\hf 13. &J.~Wichmann   
        &\in &10.03     &38.92  &o   &10.18     &38.95 \cr
\+\hf 14. &P.~Abratkiewicz       
        &\in &10.29     &39.31  &o   &10.21     &38.95 \cr
\+\hf 15. &L.~Funke              
        &o   &{\hb}9.95 &38.76  &\in &10.15     &38.87 \cr
\+\hf 16. &K.~Scott              
        &o   &10.02     &38.43  &\in &10.04     &39.06 \cr
\+\hf 17. &Y.~Fujimoto   
        &\in &{\hb}9.95 &38.46  &o   &10.11     &39.63 \cr
\+\hf 18. &D.~Cruikshank 
        &o   &10.16     &38.69  &\in &10.24     &39.10 \cr
\+\hf 19. &B.~Forslund   
        &\in &10.20     &38.95  &o   &10.28     &39.01 \cr
\+\hf 20. &E.~Flaim              
        &\in &10.23     &39.70  &o   &10.33     &39.46 \cr 
\+\hf 21. &M.~Enfeldt    
        &\in &10.12     &39.23  &o   &10.36     &39.14 \cr
\+\hf 22. &R.~Brunner    
        &\in &10.11     &39.43  &o   &10.13     &39.67 \cr
\+\hf 23. &M.~Hadschieff 
        &o   &10.33     &39.81  &\in &10.44     &39.56 \cr
\+\hf 24. &H.~Ilkka              
        &\in &10.16     &38.89  &o   &10.20     &39.01 \cr
\+\hf 25. &I.~Dolp               
        &o   &10.52     &40.74  &\in &10.50     &41.29 \cr
\+\hf 26. &A.~Golubyev   
        &\in &disq      &       &o   &{\hb}9.92 &38.20 \cr
\+\hf 27. &A.~Loef               
        &o   &10.12     &38.61  &\in &fell       \cr
\+\hf 28. &G.~Thibault           
        &o   &10.03     &38.62  &\in &10.14     &38.28  \cr
\+\hf 29. &R.~Dubreuil   
        &o   &10.19     &39.22  &\in &10.42     &39.60  \cr

\bigskip
\+ &{\bf Inzell 1991, Feb 23--24:} \cr 

\medskip        
\+\hf  1. &I.~Zhelezovsky       
        &\in &{\hb}9.93 &37.11  &o   &{\hb}9.87 &36.92 \cr
\+\hf  2. &U.-J.~Mey                    
        &\in &{\hb}9.69 &36.96  &o   &{\hb}9.65 &36.64 \cr
\+\hf  3. &T.~Kuroiwa    
        &\in &10.21     &37.31  &o   &10.01     &37.72 \cr
\+\hf  4. &D.~Jansen            
        &\in &{\hb}9.85 &37.23  &o   &{\hb}9.81 &36.94 \cr
\+\hf  5. &Yas.~Miyabe      
        &o   &10.15     &37.78  &\in &{\hb}9.80 &36.78 \cr
\+\hf  6. &Y.~Fujimoto  
        &o   &{\hb}9.80 &37.41  &\in &{\hb}9.70 &37.25 \cr
\+\hf  7. &A.~Bakhvalov 
        &o   &{\hb}9.95 &37.45  &\in &{\hb}9.89 &37.23 \cr
\+\hf  8. &N.~Gulyayev  
        &\in &10.17     &37.84  &o   &10.17     &37.74 \cr
\+\hf  9. &P.~Adeberg   
        &\in &{\hb}9.88 &37.63  &o   &{\hb}9.87 &37.51 \cr
\+\hf 10. &D.~Besteman  
        &o   &10.18     &37.87  &\in &10.11     &37.62 \cr
\+\hf 11. &Shakshakbayev        
        &o   &10.04     &38.21  &\in &{\hb}9.80 &36.95 \cr
\+\hf 12. &S.~Ireland           
        &\in &10.15     &37.85  &o   &10.09     &37.61 \cr
\+\hf 13. &S.-Y.~Chegal 
        &\in &{\hb}9.94 &37.62  &o   &{\hb}9.98 &37.40 \cr
\+\hf 14. &G.~Thibault          
        &o   &10.26     &38.63  &\in &{\hb}9.98 &37.26 \cr
\+\hf 15. &G.~v Velde 
        &o   &10.34     &38.26  &\in &10.07     &37.72 \cr
\+\hf 16. &N.~Thometz           
        &o   &10.19     &38.04  &\in &10.09     &37.90 \cr
\+\hf 17. &N.~Mills     
        &o   &10.35     &38.40  &\in &10.21     &38.19 \cr
\+\hf 18. &R.~Brunner   
        &\in &10.09     &38.30  &o   &{\hb}9.90 &38.05 \cr
\+\hf 19. &B.~Forslund  
        &o   &10.20     &38.21  &\in &10.05     &37.88 \cr
\+\hf 20. &Y.-M.~Kim            
        &\in &10.05     &37.96  &o   &{\hb}9.95 &37.91 \cr
\+\hf 21. &K.~Scott             
        &\in &10.05     &38.37  &o   &10.03     &38.10 \cr
\+\hf 22. &R.~Str\o m   
        &o   &{\hb}9.99 &38.06  &\in &10.00     &37.68 \cr
\+\hf 23. &M.~Enfeldt   
        &\in &10.23     &38.28  &o   &10.23     &38.26 \cr
\+\hf 24. &P.~Kelly             
        &\in &10.16     &38.49  &o   &10.02     &37.91 \cr
\+\hf 25. &H.~Markstr\"om       
        &\in &10.28     &38.54  &o   &10.17     &37.96 \cr
\+\hf 26. &D.~Kah               
        &o   &10.69     &39.11  &\in &10.69     &38.98 \cr
\+\hf 27. &H.~Janssen           
        &o   &10.51     &39.21  &\in &10.35     &38.72 \cr
\+\hf 28. &P.~Abratkiewicz      
        &o   &10.13     &38.94  &\in &10.09     &38.55 \cr
\+\hf 29. &H.~Ilkka             
        &\in &10.28     &38.51  &o   &10.31     &38.44 \cr
\+\hf 30. &C.~McNicoll  
        &o   &10.56     &39.49  &\in &10.58     &39.15 \cr
\+\hf 31. &V.~Gios      
        &o   &10.21     &39.13  &\in &10.14     &38.89 \cr
\+\hf 32. &I.~Dolp              
        &\in &10.07     &39.84  &o   &10.42     &39.82 \cr
\+\hf 33. &Y.~Kuroiwa   
        &\in &10.00     &37.28  &o   &{\hb}9.85 &fell  \cr
\eject
\+\hf 34. &U.~Streb             
        &\in &10.13     &37.75  &o   &10.04     &37.60 \cr
\+\hf 35. &G.~vd Brink 
        &o   &10.27     &38.82  &\in &10.24     &38.77 \cr

\bigskip
\+ &{\bf Troms\o{} 1990, Feb 24--25:} \cr 


\medskip        
\+\hf  1. &K.-T.~Bae             
        &o    &10.05    &38.16   &\in &10.15 &38.24 \cr
\+\hf  2. &A.~Bakhvalov          
        &\in &{\hb}9.86 &38.54    &o   &10.04 &38.15 \cr
\+\hf  3. &I.~Zhelezovsky        
        &\in &{\hb}9.89 &37.85   &o   &10.14 &38.78 \cr
\+\hf  4. &D.~Jansen             
        &\in &{\hb}9.64 &38.44   &o   &{\hb}9.99 &38.15 \cr
\+\hf  5. &Yas.~Miyabe   
        &\in &{\hb}9.82 &38.35   &o   &10.05 &38.07 \cr
\+\hf  6. &A.~Klimov      
        &\in &10.20     &39.19   &o   &10.25 &38.81 \cr
\+\hf  7. &H.~Moriyama   
        &o   &10.06     &39.24   &\in &10.18 &38.29 \cr
\+\hf  8. &S.~Kuroiwa    
        &o   &{\hb}9.89 &37.98   &\in &10.03 &38.31 \cr
\+\hf  9. &T.~Kuroiwa    
        &\in &10.39     &39.32   &o   &10.47 &38.75 \cr
\+\hf 10. &O.~Zinke              
        &o   &10.16     &39.24   &\in &10.46 &39.15 \cr
\+\hf 11. &A.~Loef               
        &o   &10.13     &38.77   &\in &10.39 &38.98 \cr
\+\hf 12. &D.~Besteman   
        &\in &10.19     &38.88   &o   &10.34 &38.99 \cr
\+\hf 13. &K.~Scott      
        &o   &{\hb}9.99 &39.21   &\in &10.16 &38.67 \cr
\+\hf 14. &G.~Thibault           
        &o   &{\hb}9.98 &39.00   &\in &10.17 &38.85 \cr
\+\hf 15. &R.~Dubreuil   
        &\in &10.13     &39.22   &o   &10.32 &39.94 \cr
\+\hf 16. &E.~Flaim              
        &o   &10.05     &39.55   &\in &10.42 &39.10 \cr
\+\hf 17. &N.~Thometz    
        &o   &10.04     &39.16   &\in &10.30 &39.20 \cr
\+\hf 18. &B.~K\"onig            
        &\in &10.17     &39.61   &o   &10.39 &39.36 \cr
\+\hf 19. &M.~Boelsma    
        &\in &{\hb}9.91 &39.42   &o   &10.36 &39.16 \cr
\+\hf 20. &B.~Forslund           
        &o   &10.37     &39.74   &\in &10.43 &39.19 \cr
\+\hf 21. &G.~vd Brink  
        &o   &10.36     &39.90   &\in &10.61 &39.73 \cr
\+\hf 22. &S.-Y.~Chegal          
        &o   &{\hb}9.90 &38.99   &\in &10.18 &39.19 \cr
\+\hf 23. &L.~Wei                
        &o   &10.48     &40.13   &\in &10.57 &39.55 \cr
\+\hf 24. &A.~Franzelin          
        &\in &10.58     &39.67   &o   &10.67 &40.04 \cr
\+\hf 25. &A.~Hoffmann   
        &\in &10.43     &40.22   &o   &10.45 &39.94 \cr
\+\hf 26. &U.~Pikkupeura         
        &\in &10.51     &40.32   &o   &10.76 &39.90 \cr
\+\hf 27. &R.~Str\o m    
        &o   &10.02     &39.41   &\in &10.35 &39.42 \cr
\+\hf 28. &T.~Jankowski          
        &\in &10.22     &fell    &o   &10.36 &39.75 \cr
\+\hf 29. &T.~Terpstra   
        &\in &10.16     &39.54   &o   &10.30 &fell  \cr
\+\hf 30. &U.-J.~Mey     
        &\in &{\hb}9.71 &37.16   &o   &{\hb}9.95 &37.83 \cr
\+\hf 31. &U.~Streb              
        &o   &disq      &        &\in &10.50 &39.09 \cr

\bigskip
\+ &{\bf Heerenveen 1989, Feb 25--26:} \cr 


\medskip        
\+\hf  1. &I.~Zhelezovsky       
        &\in &{\hb}9.71 &36.52  &o   &{\hb}9.80 &36.83 \cr
\+\hf  2. &U.-J.~Mey            
        &o    &{\hb}9.68 &36.63 &\in &{\hb}9.74 &36.46 \cr
\+\hf  3. &A.~Bakhvalov 
        &\in &{\hb}9.90 &36.92  &o   &{\hb}9.88 &36.82 \cr
\+\hf  4. &D.~Jansen            
        &o    &{\hb}9.71 &36.87 &\in &{\hb}9.78 &36.55 \cr
\+\hf  5. &N.~Thometz           
        &\in &{\hb}9.89 &36.93  &o   &{\hb}9.98 &37.10  \cr
\+\hf  6. &K.-T.~Bae            
        &o    &{\hb}9.89 &36.90 &\in &{\hb}9.87 &36.90 \cr
\+\hf  7. &E.~Flaim             
        &\in &{\hb}9.99 &37.21  &o   &10.06     &37.19 \cr
\+\hf  8. &G.~Thibault          
        &o    &{\hb}9.95 &37.01 &\in &10.01     &37.20 \cr
\+\hf  9. &A.~Sobeck            
        &\in &{\hb}9.91 &36.96  &o   &{\hb}9.94 &37.06 \cr
\+\hf 10. &Y.~Kuroiwa   
        &\in &{\hb}9.95 &37.21  &o   &{\hb}9.86 &37.02 \cr
\+\hf 11. &H.~Moriyama  
        &o    &{\hb}9.90 &37.05 &\in &{\hb}9.96 &37.40 \cr
\+\hf 12. &J.~Ykema             
        &\in &10.12     &37.61  &o   &10.13     &37.52 \cr
\+\hf 13. &D.~Cruikshank        
        &o    &10.03     &37.41 &\in &10.08     &37.42 \cr
\+\hf 14. &R.~Dubreuil  
        &\in &10.17     &37.43  &o   &10.12     &37.50 \cr
\+\hf 15. &A.~Loef              
        &\in &10.21     &37.69  &o   &10.13     &37.82 \cr
\+\hf 16. &M.~Hadschieff        
        &\in &10.23     &37.97  &o   &10.19     &37.85 \cr
\+\hf 17. &T.~Terpstra  
        &\in &{\hb}9.90 &37.76  &o   &10.01     &37.56 \cr
\+\hf 18. &Y.~Mitani    
        &o    &10.07     &37.43 &\in &10.19     &37.35 \cr
\+\hf 19. &B.~Forslund  
        &o    &10.23     &37.98 &\in &10.13     &37.71 \cr
\+\hf 20. &M.~Richmond  
        &o    &10.28     &38.00 &\in &10.30     &37.90 \cr
\+\hf 21. &F.~R\o nning 
        &\in &10.16     &37.68  &o   &10.17     &37.87 \cr
\+\hf 22. &B.~Jonkman   
        &o    &{\hb}9.98 &38.16 &\in &10.13     &38.02 \cr
\+\hf 23. &U.~Streb             
        &\in &10.22     &38.24  &o   &10.19     &38.11 \cr
\+\hf 24. &R.~Sighel             
        &o    &10.40     &38.51 &\in &10.32     &38.99 \cr
\+\hf 25. &B.~K\"onig           
        &\in &10.16     &38.11  &o   &10.19     &38.55 \cr
\+\hf 26. &U.~Pikkopeura        
        &o    &10.31     &38.41 &\in &10.35     &38.17 \cr
\+\hf 27. &Y.-I.~Ha             
        &o    &10.18     &38.31 &\in &10.24     &38.25 \cr
\+\hf 28. &P.~Tahmindjis        
        &o    &10.57     &39.11 &\in &10.69     &39.46 \cr
\+\hf 29. &I.~Heikkil\"a
        &o    &10.28     &39.09 &\in &10.24     &38.50 \cr
\+\hf 30. &T.~Aouyanagi 
        &\in &10.32     &37.94  &o   &10.26     &fell  \cr
\+\hf 31. &B.~Merdovic  
        &o    &10.97     &41.67 &\in &10.93     &41.08 \cr
\+\hf 32. &R.~Str\o m           
        &\in &{\hb}9.99 &fell  &o    &{\hb}9.94 &37.82 \cr
\+\hf 33. &S.~Fokitshev         
        &\in &{\hb}9.89 &37.14  &o   &{\hb}9.90 &fell  \cr
\+\hf 34. &\AA.~S\o ndr\aa l    
        &o    &10.29     &fell  &\in &10.24     &38.17 \cr
\+\hf 35. &T.~Jankowski         
        &\in &{\hb}9.99 &disq   &o   &10.00     &37.86 \cr

\bigskip
\+ &{\bf West Allis 1988, Feb 6--7:} \cr 

\medskip
\+\hf  1. &D.~Jansen            
        &o   &{\hb}9.82 &38.70  &\in &{\hb}9.72 &38.15 \cr 
\+\hf  2. &U.-J.~Mey            
        &o   &{\hb}9.86 &38.89  &\in &{\hb}9.88 &38.72 \cr 
\+\hf  3. &E.~Flaim             
        &\in &10.16     &39.17  &o   &10.10 &38.96 \cr 
\+\hf  4. &A.~Bakhvalov 
        &\in &10.00     &38.96  &o   &{\hb}9.93 &38.51 \cr 
\+\hf  5. &E.~Henriksen 
        &\in &10.23     &39.08  &o   &10.51 &38.70 \cr 
\+\hf  6. &J.~Ykema             
        &o   &10.09     &39.39  &\in &{\hb}9.99 &38.74 \cr 
\+\hf  7. &K.~Hamaya    
        &\in &10.10     &39.92  &o   &{\hb}9.99 &38.22 \cr 
\+\hf  8. &Y.~Kanehama  
        &o   &10.12     &38.88  &\in &10.15 &39.11 \cr 
\+\hf  9. &N.~Thometz           
        &\in &10.01     &39.34  &o   &10.08 &38.68 \cr 
\+\hf 10. &V.~Makovetski        
        &o   &{\hb}9.94 &38.94  &\in &{\hb}9.77 &38.83 \cr 
\+\hf 11. &Y.~Mitani    
        &o   &10.30     &39.55  &\in &10.70 &38.76 \cr 
\+\hf 12. &G.~Thibault          
        &o   &10.10     &39.59  &\in &{\hb}9.83 &38.98 \cr 
\+\hf 13. &F.~R\o nning 
        &o   &10.29     &39.57  &\in &10.11 &39.20 \cr 
\+\hf 14. &J.~Vesterlund        
        &\in &10.09     &39.70  &o   &{\hb}9.97 &38.85 \cr 
\+\hf 15. &H.~Moriyama  
        &\in &10.05     &39.38  &o   &10.08 &39.39 \cr 
\+\hf 16. &B.~Hagen             
        &o   &10.22     &39.93  &\in &10.17 &39.56 \cr 
\+\hf 17. &M.~Boelsma   
        &\in &10.05     &40.02  &o   &10.00 &40.03 \cr 
\+\hf 18. &B.~Forslund  
        &\in &10.37     &40.77  &o   &10.18 &39.83 \cr 
\+\hf 19. &A.~Loef              
        &\in &10.36     &40.11  &o   &10.30 &39.83 \cr 
\+\hf 20. &M.~Richmond   
        &o   &10.38     &40.27  &\in &10.26 &39.79 \cr 
\+\hf 21. &U.~Streb             
        &o   &10.25     &40.30  &\in &10.08 &40.01 \cr 
\+\hf 22. &U.~Pikkupeura        
        &\in &10.30     &40.13  &o   &10.34 &40.60 \cr 
\+\hf 23. &H.-P.~Oberhuber      
        &o   &10.21     &40.14  &\in &10.09 &39.85 \cr 
\+\hf 24. &A.~Hoffmann  
        &o   &10.07     &39.87  &\in &10.11 &40.41 \cr 
\+\hf 25. &M.~Vernier           
        &o   &10.40      &41.11 &\in &10.29 &40.57 \cr 
\+\hf 26. &P.~Hinni     
        &\in &10.63     &41.99  &o   &10.30 &40.65 \cr 
\+\hf 27. &B.~Repnin            
        &\in &10.10     &40.30  &o   &10.12 &39.36 \cr 
\+\hf 28. &J.~Dominik   
        &o   &10.12     &39.84  &\in &10.11 &39.87 \cr 
\+\hf 29. &G.~Boucher   
        &\in &disq      &       &o   &{\hb}9.93 &39.73 \cr 
\+\hf 30. &K.-T.~Bae            
        &\in &didn't start &    &didn't start \cr 

\bigskip
\+ &{\bf Sainte Foy 1987, Jan 31--Feb 1:} \cr 

\medskip
\+\hf  1. &A.~Kuroiwa   
        &o &10.02 &38.20        &\in  &{\hb}9.93 &37.90 \cr 
\+\hf  2. &N.~Thometz           
        &\in &{\hb}9.96  &38.53 &o  &{\hb}9.97 &37.46 \cr 
\+\hf  3. &Y.~Mitani    
        &\in &10.14 &38.71      &o  &10.15 &37.93 \cr 
\+\hf  4. &U.-J.~Mey            
        &\in &{\hb}9.73  &39.23 &o  &{\hb}9.87 &38.02 \cr 
\+\hf  5. &S.~Fokitshev         
        &o &{\hb}9.77  &37.99   &\in  &{\hb}9.89 &37.82 \cr 
\+\hf  6. &G.~Kuiper            
        &\in &10.26 &38.70      &o &10.05 &38.43 \cr 
\+\hf  7. &A.~Bakhvalov 
        &\in &10.05 &38.98      &o &10.06 &38.53 \cr 
\+\hf  8. &I.~Zhelezovsky       
        &\in &10.29 &40.06      &o &10.25 &38.37 \cr 
\+\hf  9. &E.~Henriksen 
        &o &10.63 &39.27        &\in &10.41 &38.54 \cr 
\+\hf 10. &U.~Streb             
        &o &10.34 &39.32        &\in &10.28 &39.10 \cr 
\+\hf 11. &M.~Hirose    
        &\in &10.15 &39.28      &o  &{\hb}9.98 &38.16 \cr 
\+\hf 12. &D.~Jansen            
        &o &{\hb}9.91  &39.19   &\in  &{\hb}9.94 &38.16 \cr 
\+\hf 13. &F.~R\o nning 
        &\in &10.20 &39.16      &o &10.27 &38.65 \cr 
\+\hf 14. &Y.~Kanehama  
        &o &10.33 &39.82        &\in &10.30 &38.73 \cr 
\+\hf 15. &G.~Boucher   
        &o &10.03 &39.14        &\in &10.10 &39.14 \cr 
\+\hf 16. &K.-T.~Bae            
        &o &10.25 &39.82        &\in &10.30 &39.03 \cr 
\+\hf 17. &K.~Kondratsyev
        &\in &10.12 &39.37      &o &10.15 &38.84 \cr 
\+\hf 18. &J.~Ykema             
        &o &10.10 &39.42        &\in &10.16 &38.50 \cr 
\+\hf 19. &A.~Hoffmann  
        &\in &10.22 &40.06      &o &10.26 &38.83 \cr 
\+\hf 20. &J.~Vesterlund        
        &\in &10.19 &39.00      &o &10.20 &38.75 \cr 
\+\hf 21. &G.~Thibault          
        &\in &10.33 &39.29      &o &10.19 &38.99 \cr 
\+\hf 22. &B.~Jonkman           
        &o &10.29 &39.76        &\in &10.30 &39.26 \cr 
\+\hf 23. &M.~Hadschieff        
        &\in &10.44 &40.55      &o &10.51 &39.35 \cr 
\+\hf 24. &H.-P.~Oberhuber 
        &\in &10.32 &39.46      &o &10.30 &38.81 \cr 
\+\hf 25. &G.~Johansson         
        &o &10.37 &39.53        &\in &10.23 &39.26 \cr 
\+\hf 26. &B.~Hagen             
        &\in &10.36 &40.03      &o &10.49 &39.31 \cr 
\+\hf 27. &M.~Richmond  
        &\in &10.64 &40.49      &o &10.35 &39.24 \cr 
\+\hf 28. &D.~Turcotte  
        &o &10.03 &39.76        &\in &10.13 &39.26 \cr 
\+\hf 29. &Y.-S.~Ra             
        &o &10.28 &40.17        &\in &10.33 &39.33 \cr 
\+\hf 30. &P.~Hinni             
        &\in &10.44 &40.90      &o &10.65 &40.43 \cr 
\+\hf 31. &M.~Vernier   
        &o &10.44 &41.00        &\in &10.52 &40.57 \cr  
\+\hf 32. &I.~Bredeli   
        &o &10.36 &39.08        &\in &10.31 &39.15 \cr 
\+\hf 33. &J.~Dominik   
        &o &10.38 &39.77        &\in &10.29 &39.47 \cr 
\+\hf 34. &P.~Tahmindjis        
        &o &11.12 &41.87        &\in &11.22 &fell \cr    

\bigskip
\+ &{\bf Karuizawa 1986, Feb 22--23:} \cr 

\medskip
\+\hf  1. &I.~Zhelezovsky       
        &\in &{\hb}9.95 &37.35  &o  &{\hb}9.92 &37.26 \cr
\+\hf  2. &D.~Jansen            
        &o   &{\hb}9.78 &37.38  &\in &{\hb}9.62 &36.84 \cr
\+\hf  3. &A.~Kuroiwa   
        &o   &{\hb}9.90 &37.13  &\in &{\hb}9.95 &37.32 \cr
\+\hf  4. &N.~Thometz           
        &o   &10.05 &37.70      &\in &{\hb}9.95 &37.78 \cr
\+\hf  5. &U.-J.~Mey            
        &\in &{\hb}9.83 &37.48  &o   &{\hb}9.74 &37.26 \cr
\+\hf  6. &Y.~Mitani    
        &\in &{\hb}9.96 &37.62  &o   &10.05 &37.97 \cr
\+\hf  7. &S.~Fokitshev 
        &\in &{\hb}9.68 &37.64  &o   &{\hb}9.76 &37.37 \cr
\+\hf  8. &G.~Kuiper            
        &\in &{\hb}9.89 &37.91  &o   &{\hb}9.85 &37.69 \cr
\+\hf  9. &M.~Hirose    
        &\in &10.35 &38.14      &o &10.05 &37.84 \cr
\+\hf 10. &K.~Hamaya    
        &\in &10.21 &38.07      &o  &{\hb}9.94 &38.29 \cr
\+\hf 11. &K.~Kondratsyev       
        &o   &{\hb}9.88 &37.91  &\in  &{\hb}9.78 &37.92 \cr
\+\hf 12. &G.~Thibault          
        &\in &10.45 &38.26      &o &10.09 &38.12 \cr
\+\hf 13. &G.~Boucher   
        &o   &{\hb}9.95 &38.34  &\in  &{\hb}9.92 &39.20 \cr
\+\hf 14. &F.~R\o nning 
        &o   &10.00 &37.70      &\in &10.12 &38.20 \cr
\+\hf 15. &M.~Richmond  
        &o   &10.41 &38.63      &\in &10.26 &38.42 \cr
\+\hf 16. &H.~Vergeer           
        &\in &10.31 &38.89      &o &10.31 &38.40 \cr
\+\hf 17. &Y.-S.~Ra             
        &o 1 &10.27 &38.95      &\in &10.21 &38.31 \cr
\+\hf 18. &I.~Bredeli   
        &\in &10.06 &38.22      &o &10.42 &38.46 \cr
\+\hf 19. &U.~Streb             
        &o   &10.33 &38.99      &\in &10.15 &38.28 \cr
\+\hf 20. &M.~Hadschieff        
        &o   &10.57 &39.27      &\in &10.27 &38.86 \cr
\+\hf 21. &E.~Henriksen 
        &o   &10.69 &39.03      &\in &10.44 &38.59 \cr
\+\hf 22. &J.~Dominik   
        &\in &10.31 &38.71      &o &10.09 &38.61 \cr
\+\hf 23. &G.~Johansson 
        &\in &10.30 &38.56      &o &10.08 &38.28 \cr
\+\hf 24. &R.~Rzadki    
        &o   &10.38 &39.34      &\in &10.14 &38.50 \cr
\+\hf 25. &J.~Vesterlund        
        &o   &10.21 &38.72      &\in &10.28 &38.54 \cr
\+\hf 26. &K.A.~Engelstad       
        &\in &10.34 &39.10      &o &10.37 &39.17 \cr
\+\hf 27. &H.~Janssen           
        &o   &10.84 &39.62      &\in &10.63 &39.57 \cr
\+\hf 28. &C.~Nicoleau  
        &\in &10.37 &38.99      &o  &{\hb}9.99 &38.76 \cr
\+\hf 29. &W.~J\"ager   
        &\in &10.84 &40.25      &o &10.61 &40.18 \cr
\+\hf 30. &H.-P.~Oberhuber      
        &o   &10.75 &39.70      &\in &10.36 &39.28 \cr
\+\hf 31. &J.~Ykema             
        &\in &10.11 &38.09      &o &10.15 &37.89 \cr
\+\hf 32. &B.~Repnin    
        &o   &10.15 &38.66      &\in &10.06 &fell \cr
\+\hf 33. &T.-S.~Yen      
        &withdrawn & &          &o &12.46 &50.95 \cr

\bigskip
\+ &{\bf Heerenveen 1985, Feb 23--24:} \cr 

\medskip
\+\hf  1. &I.~Zhelezovsky       
        &\in &{\hb}9.98 &37.91  &o  &{\hb}9.92 &37.91 \cr 
\+\hf  2. &G.~Boucher   
        &\in &10.04 &38.45      &o &10.07 &38.00 \cr
\+\hf  3. &D.~Jansen            
        &o   &{\hb}9.96 &37.94  &\in &10.12 &38.22 \cr
\+\hf  4. &N.~Thometz           
        &\in &10.23 &38.62      &o &10.25 &38.52 \cr
\+\hf  5. &O.~Boshev            
        &o   &10.30 &38.73      &\in &10.41 &38.81 \cr
\+\hf  6. &U.-J.~Mey            
        &\in &{\hb}9.93 &38.33  &o &10.14 &38.79 \cr
\+\hf  7. &H.~Vergeer           
        &o   &10.25 &38.66      &\in &10.32 &38.91 \cr
\+\hf  8. &H.~vd Duim   
        &\in &10.17 &39.07      &o &10.51 &39.32 \cr
\+\hf  9. &S.~Fokitshev 
        &\in &{\hb}9.86 &38.45  &o &10.01 &38.62 \cr
\+\hf 10. &K.A.~Engelstad       
        &o   &10.30 &39.69      &\in &10.37 &39.12 \cr
\+\hf 11. &Y.~Mitani    
        &\in &10.21 &38.77      &o &10.23 &38.89 \cr
\+\hf 12. &K.~Hamaya    
        &o   &10.22 &39.19      &\in &10.35 &39.03 \cr
\+\hf 13. &A.~Hoffmann  
        &o   &10.26 &39.31      &\in &10.44 &39.40 \cr
\+\hf 14. &R.~Rzadki    
        &o   &10.37 &39.50      &\in &10.50 &39.12 \cr
\+\hf 15. &G.~Kuiper            
        &o   &10.36 &39.24      &\in &10.28 &38.87 \cr
\+\hf 16. &Y.~Kanehama  
        &\in &10.39 &39.31      &o &10.35 &39.23 \cr
\+\hf 17. &F.~R\o nning 
        &o   &10.23 &38.61      &\in &10.42 &38.90 \cr
\+\hf 18. &U.~Streb             
        &\in &10.40 &39.20      &o &10.35 &39.22 \cr
\+\hf 19. &J.~Vesterlund        
        &o   &10.24 &39.16      &\in &10.41 &38.88 \cr
\+\hf 20. &M.~Hadschieff        
        &\in &10.58 &40.11      &o &10.75 &39.81 \cr
\+\hf 21. &A.~Kuroiwa   
        &o   &10.11 &38.91      &\in &10.25 &38.94 \cr
\+\hf 22. &M.~Jansen            
        &o   &10.27 &39.26      &\in &10.43 &39.41 \cr
\+\hf 23. &K.-T.~Bae            
        &\in &10.44 &40.32      &o &10.47 &38.92 \cr
\+\hf 24. &B.~Lamarche  
        &o   &10.52 &39.89      &\in &10.65 &40.00 \cr
\+\hf 25. &B.~Hagen             
        &o   &10.46 &39.61      &\in &10.65 &39.94 \cr
\+\hf 26. &U.~Pikkupeura        
        &\in &10.86 &40.08      &o &10.71 &39.91 \cr
\+\hf 27. &D.~Gagnon            
        &\in &10.30 &39.59      &o &10.40 &39.60 \cr
\+\hf 28. &V.~Finsand   
        &\in &10.70 &40.43      &o &10.76 &40.23 \cr
\+\hf 29. &G.~Johansson 
        &\in &10.26 &39.77      &o &10.35 &39.50 \cr
\+\hf 30. &G.~Paganin   
        &o   &10.67 &41.13      &\in &10.63 &39.93 \cr
\+\hf 31. &E.~Henriksen 
        &o   &10.34 &38.63      &\in &10.45 &38.06 \cr
\+\hf 32. &A.~Bakhvalov 
        &o   &10.24 &fell       &\in &10.39 &39.14 \cr
\+\hf 33. &U.~Gebauer           
        &\in &10.34 &fell       &o &10.31 &39.41 \cr
\+\hf 34. &M.~Vernier   
        &\in &10.41 &40.45      &o &didn't start \cr

\bigskip
\+ &{\bf Trondheim 1984, March 3--4:} \cr 

\medskip
\+\hf  1. &G.~Boucher   
        &\in &{\hb}9.98 &38.21  &o  &{\hb}9.97 &38.02 \cr 
\+\hf  2. &S.~Khlebnikov        
        &\in &10.28 &38.24      &o &10.22 &37.98 \cr
\+\hf  3. &K.A.~Engelstad       
        &o   &10.22 &38.69      &\in &10.24 &38.14 \cr
\+\hf  4. &N.~Thometz           
        &o   &10.27 &38.41      &\in &10.31 &38.12 \cr
\+\hf  5. &S.~Fokitshev 
        &\in &{\hb}9.85 &37.92  &o  &{\hb}9.90 &37.74 \cr
\+\hf  6. &A.~Danilin   
        &\in &10.08 &38.52      &o  &{\hb}9.94 &38.03 \cr
\+\hf  7. &D.~Jansen            
        &\in &10.27 &38.63      &o &10.18 &38.06 \cr
\+\hf  8. &A.~Kuroiwa   
        &o   &10.10 &38.28      &\in &10.13 &37.90 \cr
\+\hf  9. &K.~Hamaya    
        &\in &10.22 &38.72      &o &10.34 &38.72 \cr
\+\hf 10. &U.-J.~Mey            
        &o   &10.09 &38.60      &\in &10.08 &38.35 \cr
\+\hf 11. &H.~Vergeer           
        &o   &10.37 &39.12      &\in &10.31 &38.51 \cr
\+\hf 12. &J.~Thibault  
        &o   &10.11 &39.07      &\in &10.02 &38.40 \cr
\+\hf 13. &F.~R\o nning 
        &\in &10.34 &38.90      &o &10.40 &38.71 \cr
\+\hf 14. &J.~Vesterlund        
        &o   &10.20 &38.77      &\in &10.29 &38.69 \cr
\+\hf 15. &R.~Falk-Larssen      
        &o   &10.34 &39.21      &\in &10.53 &39.07 \cr
\+\hf 16. &Y.~Suzuki    
        &\in &10.08 &38.76      &o &10.13 &38.18 \cr
\+\hf 17. &D.~Immerfall 
        &o   &10.35 &38.97      &\in &10.52 &38.76 \cr
\+\hf 18. &J.~Ykema             
        &\in &10.24 &38.89      &o &10.27 &39.18 \cr
\+\hf 19. &U.~Pikkupeura     
        &o   &10.43 &39.71      &\in &10.49 &38.99 \cr
\+\hf 20. &F.~Gawenus   
        &\in &10.59 &39.36      &o &10.55 &39.19 \cr
\+\hf 21. &R.~Ket               
        &o   &10.51 &39.43      &\in &10.54 &38.96 \cr
\+\hf 22. &Y.~Kitazawa  
        &o   &10.23 &38.89      &\in &10.20 &38.19 \cr
\+\hf 23. &J.-\AA.~Carlberg     
        &\in &10.29 &39.06      &o &10.36 &39.07 \cr
\+\hf 24. &G.~Johansson 
        &\in &10.43 &39.51      &o &10.27 &38.96 \cr
\+\hf 25. &G.~Kuiper            
        &o   &10.32 &40.01      &\in &10.38 &38.84 \cr
\+\hf 26. &M.~Vernier   
        &\in &10.50 &40.13      &o &10.62 &39.90 \cr
\+\hf 27. &D.~Turcotte  
        &o   &10.27 &39.82      &\in &10.39 &39.54 \cr
\+\hf 28. &E.~Henriksen         
        &\in &didn't start &    &o &10.69 &38.55 \cr

\bye